\documentclass[11pt,a4paper]{article}
\usepackage{times,latexsym}
\usepackage{url}
\usepackage{epsfig}
\usepackage[table, x11names]{xcolor}
\usepackage{array, booktabs, boldline} %
\usepackage{cellspace}
\usepackage{breakurl}
\usepackage{array}
\usepackage{multirow}
\newcommand{\tabincell}[2]{\begin{tabular}{@{}#1@{}}#2\end{tabular}}
\usepackage[flushleft]{threeparttable}
\usepackage[T1]{fontenc}

\usepackage[acceptedWithA]{tacl2021v1}

\usepackage{xspace,mfirstuc,tabulary}

\newif\iftaclinstructions
\taclinstructionsfalse 
\iftaclinstructions
\renewcommand{\confidential}{}
\renewcommand{\anonsubtext}{(No author info supplied here, for consistency with
TACL-submission anonymization requirements)}
\newcommand{\instr}
\fi

\iftaclpubformat 

\else

\fi

\title{Pre-train, Prompt and Recommendation: A Comprehensive Survey of Language Modelling Paradigm Adaptations in Recommender Systems}

 \author{
   Peng Liu\Thanks{Equal contribution.}
   \and
   Lemei Zhang$^{*}$ 
   \and
   Jon Atle Gulla
   \\
   \ \\
   Department of Computer Science
   \\
   Norwegian University of Science and Technology
   \\
   \texttt{\{peng.liu, lemei.zhang, jon.atle.gulla\}@ntnu.no}
 }

\date{}

\begin{document}
\maketitle
\begin{abstract}
  The emergence of Pre-trained Language Models (PLMs) has achieved tremendous success in the field of Natural Language Processing (NLP) by learning universal representations on large corpora in a self-supervised manner. The pre-trained models and the learned representations can be beneficial to a series of downstream NLP tasks. This training paradigm has recently been adapted to the recommendation domain and is considered a promising approach by both academia and industry. In this paper, we systematically investigate how to extract and transfer knowledge from pre-trained models learned by different PLM-related training paradigms to improve recommendation performance from various perspectives, such as generality, sparsity, efficiency and effectiveness. Specifically, we propose a comprehensive taxonomy to divide existing PLM-based recommender systems w.r.t. their training strategies and objectives. Then, we analyze and summarize the connection between PLM-based training paradigms and different input data types for recommender systems. Finally, we elaborate on open issues and future research directions in this vibrant field.
\end{abstract}

\section{Introduction}
As an important part of the online environment, Recommender Systems (RSs) play a key role in discovering users' interests and alleviating information overload in their decision-making process. Recent years have witnessed tremendous success in recommender systems empowered by deep neural architectures and increasingly improved computing infrastructures. However, deep recommendation models are inherently data-hungry with an enormous amount of parameters to learn, which are likely to overfit and fail to generalize well in practice when their training data (i.e., user-item interactions) are insufficient. Such scenarios widely exist in practical RSs when a large number of new users join in but have fewer interactions. Consequently, the data sparsity issue becomes a major performance bottleneck of the current deep recommendation models.

With the thriving of pre-training in NLP \citep{qiu2020pre}, many language models have been pre-trained on large-scale unsupervised corpora and then fine-tuned in various downstream supervised tasks to achieve state-of-the-art results, such as GPT \citep{brown2020language}, and BERT \citep{devlin2019bert}. One of the advantages of this pre-training and fine-tuning paradigm is that it can extract informative and transferrable knowledge from abundant unlabelled data through self-supervision tasks such as masked LM \citep{devlin2019bert}, which will benefit downstream tasks when the labelled data for these tasks is insufficient and avoid training a new model from scratch. A recently proposed paradigm, prompt learning \citep{liu2023pre}, further unifies the use of pre-trained language models (PLMs) on different tasks in a simple yet flexible manner. In general, prompt learning relies on a suite of appropriate prompts, either hard text templates \citep{brown2020language}, or soft continuous embeddings \citep{qin2021learning}, to reformulate the downstream tasks as the pre-training task. The advantage of this paradigm lies in two aspects:  (1) It bridges the gap between pre-training and downstream objectives, allowing better utilization of the rich knowledge in pre-trained models. This advantage will be multiplied when very little downstream data is available. (2) Only a small set of parameters are needed to tune for prompt engineering, which is more efficient. 

Motivated by the remarkable effectiveness of the aforementioned paradigms in solving data sparsity and efficiency issues, adapting language modelling paradigms for recommendation is seen as a promising direction in both academia and industry, which has greatly advanced the state-of-the-art in RSs. Although there have been several surveys on pre-training paradigms in the fields of CV \citep{ijcai2022p773}, NLP \citep{liu2023pre} and graph learning \citep{liu2022graphsurvey}, only a handful of literature reviews are relevant to RSs. \citet{zeng2021knowledge} summarizes some research on the pre-training of recommendation models and discusses knowledge transfer methods between different domains. But it only covers a small number of BERT-like works and does not go deep into the training details of pre-trained recommendation models. \citet{yu2022self} give a brief overview of the advances of self-supervised learning in RSs. However, its focus is on a purely self-supervised recommendation setting, which means the supervision signals used to train the model are semi-automatically generated from the raw data itself. While our work does not strictly focus on the self-supervised training strategies but also incorporates the adaptation and exploration of supervised signals and data augmentation techniques in the pre-training, fine-tuning and prompting process for various recommendation purposes. Furthermore, none of them systematically analyzed the relationship between different data types and training paradigm choices in RSs. To the best of our knowledge, our survey is the first work that presents an up-to-date and comprehensive review of \textbf{L}anguage \textbf{M}odelling Paradigm Adaptations for \textbf{R}ecommender \textbf{S}ystems (LMRS)\footnote{It is worth noting that most of the existing literature reviews on pre-trained models focus on the architecture of large-scale language models (such as Bert, T5, UniLMv2, etc.), while our survey mainly discusses training paradigms, which are not limited to pre-trained language model architectures. It can also be other neural networks, such as CNN \citep{chen2021user}, and GCN \citep{liu2022graph}.}. The main contributions of this paper are summerized as follows:

\begin{itemize}
\item We survey the current state of PLM-based recommendation from perspectives of training strategy, learning objective and related data types, and provide the first systematic survey, to the best of our knowledge, in this nascent and rapidly developing field. 

\item We comprehensively review existing research works on adapting language modelling paradigms to recommendation tasks by systematically categorizing them from two perspectives: pre-training \& fine-tuning and prompting. For each category, several subcategories are provided and explained along with their concepts, formulations, involved methods, and their training and inferencing process for recommendations.  

\item We shed light on limitations and possible future research directions to help beginners and practitioners interested in this field learn more effectively with the shared integrated resources.  
\end{itemize}

\section{Generic Architecture of LMRS}
\begin{figure}
\centering
   \epsfig{figure=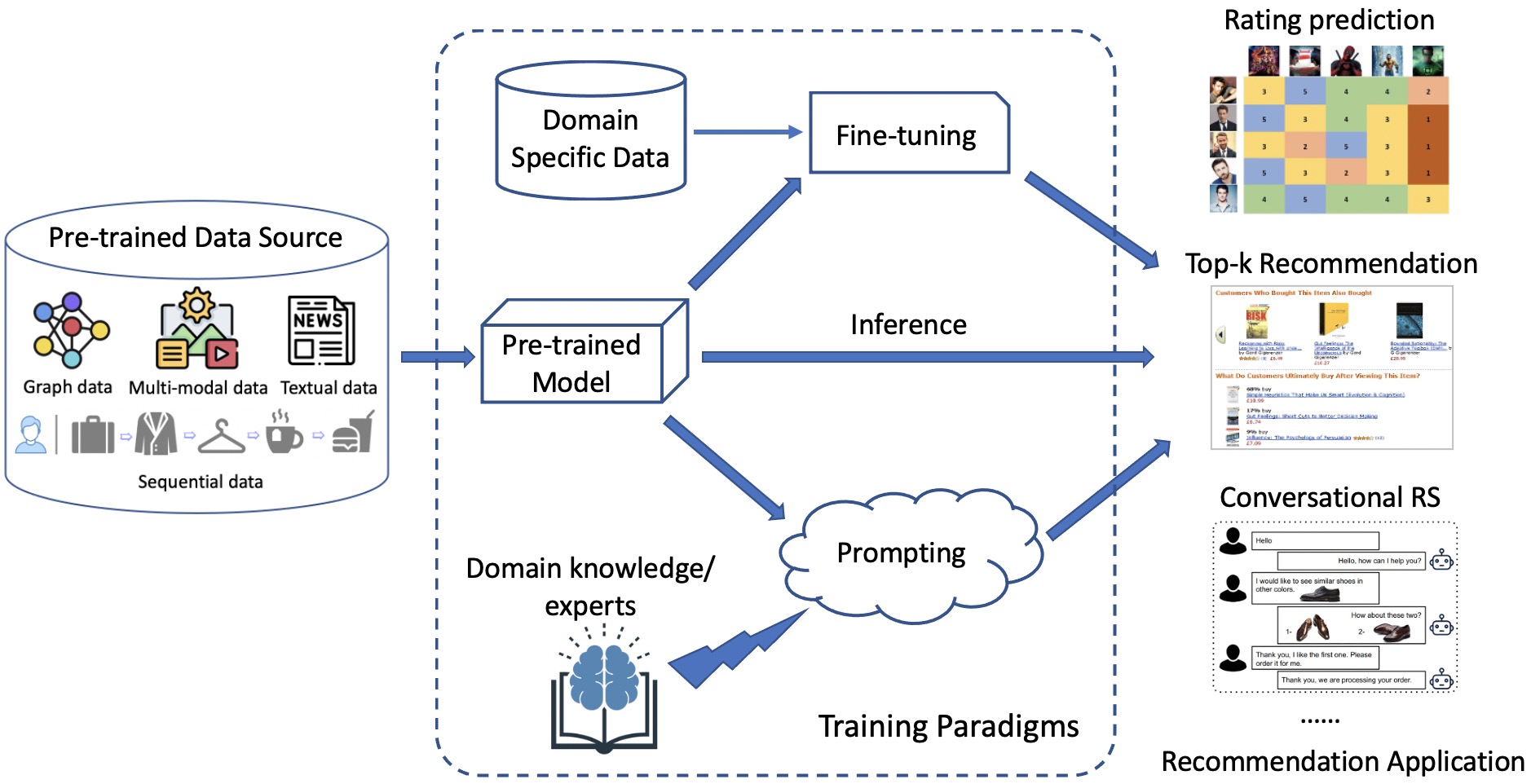,height=116pt,width=232pt}
\caption{A generic architecture of language modelling paradigm for recommendation purpose.} \label{fig:architecture}
\end{figure}

LMRS provides a new way to conquer the data sparsity problem via knowledge transfer from Pre-trained models (PTMs). Figure \ref{fig:architecture} shows a high-level overview of the LMRS, highlighting the data input, pre-training, fine-tuning/prompting and inference stages for various recommendation tasks. In general, the types of input data objects can be relevant w.r.t. both the training and inference stages. After preprocessing the input into desired forms such as graphs, ordered sequences, or aligned text-image pairs, the training process takes in the preprocessed data and performs either ``\textit{pre-train, fine-tune}" or ``\textit{pre-train, prompt}" flow. If the inference is solely based upon the pre-trained model, it can be seen as an end-to-end approach leveraging LM-based learning objectives. The trained model can then be used to infer different recommendation tasks.

\section{Data Types}
Encoding input data as embeddings is usually the first step in recommendations. However, the input for recommender systems is more diverse than most NLP tasks, and therefore, encoding techniques and processes may need to be adjusted to align with different input types. \textbf{Textual data} as a powerful medium of spreading and transmitting knowledge are commonly used as input for modelling user preferences. Examples of textual data include reviews, comments, summaries, news, conversations and codes. Note that we also consider item metadata and user profiles as a kind of textual data for simplicity. \textbf{Sequential data}, such as user-item interactions strictly arranged chronologically or in a specific order, are used as sequential input for sequential and session-based recommender systems. \textbf{Graphs}, usually containing different semantic information from other types of data inputs such as the user-user social graph or heterogeneous knowledge graph, are also commonly used to extract structural knowledge to improve recommendation performance. The diversity of online environments promotes the generation of massive multimedia content, which has been shown to improve recommendation performance in numerous research works. Therefore, \textbf{Multi-modal data} such as images, videos and audios can also be importance sources for LMRS. Multi-modal data plays a crucial role in recommendation systems. However, the utilization of multi-modal data in LMRS papers is scarce, possibly due to the absence of accessible datasets. A few scholars have gathered their individual datasets to facilitate text-video-audio tri-modal music recommendations \citep{long2023multimodal} or to establish benchmarks for shopping scenarios \citep{long2023multimodal}.

\begin{figure*}
\centering
   \epsfig{figure=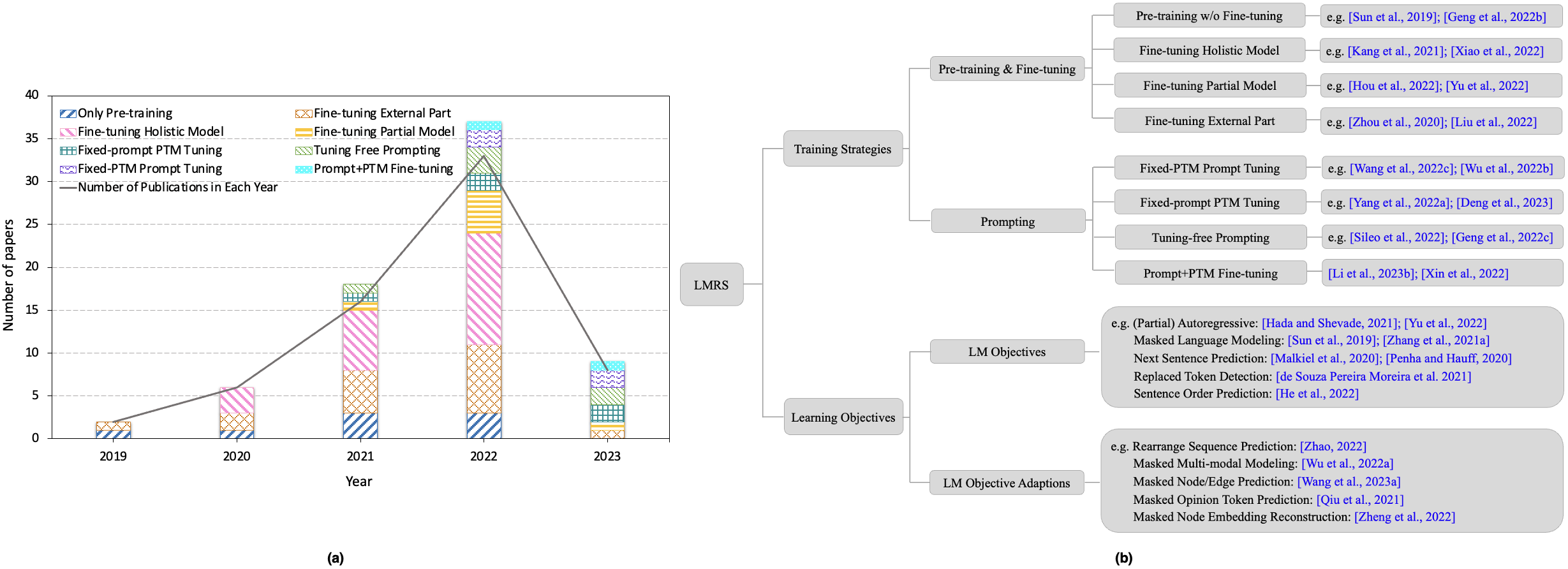,height=173pt,width=480pt}
\caption{LMRS structure with representatives and statistics on different training strategies and the total number of publications per year.} \label{fig:technology}
\end{figure*}

\section{Training Strategies of LMRS}
Given the significant impact that PLMs have had on NLP tasks in the pre-train and fine-tune paradigm, there has been a surge recently in adapting such paradigms to multiple recommendation tasks. As illustrated in Figure 1, there are mainly two classes regarding different training paradigms: pre-train, fine-tune paradigm and prompt learning paradigm. Each class is further classified into subclasses regarding different training efforts on different parts of the recommendation model. This section will go through various training strategies w.r.t. specific recommendation purposes. Figure \ref{fig:technology}(a) presents the statistics of recent publications of LMRSs grouped by different training strategies and the total number of published research works each year. Figure \ref{fig:technology}(b) shows the taxonomy and some corresponding representative LMRSs. 

\subsection{Pre-train, fine-tune paradigm for RS}
The ``\textit{pre-train, fine-tune}" paradigm attracts increasing attention from researchers in the recommendation field due to several advantages: 1) Pre-training provides a better model initialization, which usually leads to better generalization on different downstream recommendation tasks, improves recommendation performance from various perspectives, and speeds up convergence on the fine-tuning stage; 2) Pre-training on huge source corpus can learn universal knowledge which can be beneficial for the downstream recommenders; 3) Pre-training can be regarded as a kind of regularization to avoid overfitting on low-resource, and small datasets \citep{erhan2010does}. 

\noindent\textbf{Pre-train}
This training strategy can be seen as traditional end-to-end training with domain input. Differently, we only focus on research works adapting LM-based learning objectives into the training phase. Many typical LM-based RSs fall into this category, such as BERT4Rec \citep{sun2019bert4rec}, which models sequential user behaviour with a bidirectional self-attention network through Cloze task, and Transformers4Rec \citep{de2021transformers4rec} which adopts a haggingface transformer-based architecture as the base model for next-item prediction and explores four different LM tasks, namely Causal LM, MLM, Permutation LM, and Replacement Token Detection during training. These two models laid the foundation for LM-based recommender systems and have become popular baselines for their successors.

\noindent\textbf{Pre-train, fine-tune holistic model}
Under this category, the model is pre-trained and fine-tuned with different data sources, and the fine-tuning process will go through adjusting the whole model parameters. The learning objectives can also vary between the pre-training and fine-tuning stages. Pre-training and fine-tuning with different domains of data sources, also called cross-domain recommendation, can refer to the works of \citet{kang2021apirecx} and \citet{qiu2021u}. \citet{kang2021apirecx} pre-trained a GPT model using segmented source API code and fine-tuned it with API code snippets from another library for cross-library recommendation. \citet{wang2022recindial} fine-tuned the pre-trained DialoGPT model on domain-specific datasets for conversational recommendation together with an R-GCN model to inject knowledge from DBpedia to enhance recommendation performance. \citet{xiao2022training} fine-tuned the PTM to learn news embedding together with a user embedding part in an auto-regressive manner for news recommendation. They also explored different fine-tuning strategies like tuning part of the PTM and tuning the last layer of the PTM but empirically found fine-tuning the whole model resulted in better performance, which gives us an insight into balancing the recommendation accuracy and training efficiency.

\noindent\textbf{Pre-train, fine-tune partial model}
Since fine-tuning the whole model is usually time-consuming and less flexible, many LMRSs choose to fine-tune partial parameters of the model to achieve a balance between training overhead and recommendation performance \citep{hou2022towards,yu2021tiny,wu2022mm}. For instance, to deal with the domain bias problem that BERT induces a non-smooth anisotropic semantic space for general texts resulting in a large language gap for texts from different domains of items, \citet{hou2022towards} applied a linear transformation layer to transform BERT representations of items from different domains followed by an adaptive combination strategy to derive a universal item representation. Meanwhile, considering the seesaw phenomenon that learning from multiple domain-specific behavioural patterns can be a conflict, they proposed sequence-item and sequence-sequence contrastive tasks for multi-task learning during the pre-training stage. They found only fine-tuning a small proportion of model parameters could quickly adapt the model to unseen domains with cold-start or new items.

\noindent\textbf{Pre-train, fine-tune extra part of the model} 
With the increase in the depth of PTMs, the representation captured by them makes the downstream recommendation easier. Apart from the aforementioned two fine-tuning strategies, some works leverage a task-specific layer on top of the PTMs for recommendation tasks. Fine-tuning only goes through such extra parts of the PTMs by optimizing the parameters of the task-specific layer. \citet{ijcai2019p825} pre-trained a GPT and a BERT model to learn patient visit embeddings, which were then used as input to fine-tune the extra prediction layer for medication recommendation. Another approach is to use the PTM to initialize a new model with a similar architecture in the fine-tuning stage, and the fine-tuned model is used for recommendations. In \citet{zhou2020s3}, a bidirectional Transformer-based model was first pre-trained on four different self-supervised learning objectives (associated attribute prediction, masked item prediction, masked attribute prediction and segment prediction) to learn item embeddings. Then, the learned model parameters were adopted to initialize a unidirectional Transformer-based model for fine-tuning with pairwise rank loss for recommendation. In \citep{McKee_2023_CVPR}, the authors leveraged the pre-trained BLOOM-176B to generate natural languages descriptions of music given a set of music tags. Subsequently, two distinct pre-trained models, namely CLIP and the D2T pipeline, were employed to initialize textual, video, and audio representations of the provided music content. Following this, a transformer-based architecture model was fine-tuned for multi-modal music recommendation.

\subsection{Prompting paradigm for RSs}
Instead of adapting PLMs to different downstream recommendation tasks by designing specific objective functions, a rising trend in recent years is to use the ``\textit{pre-train, prompt, and inference}" paradigm to reformulate downstream recommendations through hard/soft prompts. In this paradigm, fine-tuning can be avoided, and the pre-trained model itself can be directly employed to predict item ratings, generate top-k item ranking lists, make conversations, recommend similar libraries for programmers while coding, or even output subtasks related to recommendation targets such as explanations \citep{TOIS23-PEPLER}. Prompt learning breaks through the problem of data constraints and bridges the gap of objective forms between pre-training and fine-tuning.

\noindent\textbf{Fixed-PTM prompt tuning}
Prompt-tuning only requires tuning a small set of parameters for the prompts and labels, which is especially efficient for few-shot recommendation tasks. Despite the promising results achieved through constructing prompt information without significantly changing the structure and parameters of PTMs, it also calls for the necessity of choosing the most appropriate prompt template and verbalizer, which can greatly impact recommendation performance. Prompt tuning can be both in the form of discrete textual templates \citep{penha2020does}, which are more human-readable, and soft continuous vectors \citep{wang2022towards,wu2022personalized}. For instance, \citet{penha2020does} manually designed several prompt templates to test the performance of movie/book recommendations on a pre-trained BERT model with a similarity measure. \citet{wu2022personalized} proposed a personalized prompt generator tuned to generate a soft prompt as a prefix before the user behaviour sequence for sequential recommendation.

\noindent\textbf{Fixed-prompt PTM tuning}
Fixed-prompt PTM tuning tunes the parameters of PTMs similarly to the ``pre-train, fine-tune" strategy but additionally uses prompts with fixed parameters to steer the recommendation task. One natural way is to use artificially designed discrete prompt to specify recommendation items. For instance, \citet{zhang2021language} designed a prompt ``A user watched item A, item B, and item C. Now the user may want to watch () " to reformulate the recommendation as a multi-token cloze task during fine-tuning of the LM-based PTM. The prompts can also be one or several tokens/words to seamlessly shift/lead the conversations from various tasks. \citet{deng2022unified} concatenate input sequences with special designed prompts, such as \textit{[goal]}, \textit{[topic]}, \textit{[item]}, and \textit{[system]}, to indicate different tasks: goal planning, topic prediction, item recommendation, and response generation in conversations. The model is trained using a multi-task learning scheme, and the parameters of the PTM are optimized with the same objective. \citet{yang2022improving} designed a \textit{[REC]} token as a prompt to indicate the start of the recommendation process and to summarize the dialogue context for the conversational recommendation. 

\noindent\textbf{Tuning-free prompting}
This training strategy can be referred to as \textit{zero-shot recommendations}, which directly generate recommendations or/and related subtasks without changing the parameters of the PTMs but based only on the input prompts. Zero-shot recommendation has been shown to be effective in dealing with new users/items in one domain or cross-domain settings \citep{sileo2022zero,geng2022recommendation}, compared to state-of-the-art baselines. Specifically, \citet{geng2022recommendation} learned multiple tasks, such as sequential recommendation, rating prediction, explanation generation, review summarization and direct recommendation, in a unified way with the same Negative Log-likelihood (NLL) training objectives during pre-training. At the inference stage, a series of carefully designed discrete textual template prompts were taken as input, including prompts for recommending items in the new domain (not appearing in the pre-training phase), and the trained model outputs the preferable results without a fine-tuning stage. The reason for the effectiveness of zero-shot recommendation is that the training data and pre-training tasks are able to distil rich knowledge of semantics and correlations from diverse modalities into user and item tokens, which can comprehend user preference behaviours w.r.t. item characteristics \citep{geng2022recommendation}. Building upon this research, \citet{geng2023vip5} extended their efforts to train an adapter for diverse multimodal assignments, including sequential recommendations, direct recommendations, and the generation of explanations. In particular, they utilized the pre-trained CLIP component to convert images into image tokens. These tokens were added to the textual tokens of an item to create a personalized multimodal soft prompt. This combined prompt was then used as input to fine-tune the adapter in an autoregressive manner.

\noindent\textbf{Prompt+PTM tuning}
In this setting, the parameters include two parts: prompt-relevant parameters and model parameters. The tuning phase involves optimizing all parameters for specific recommendation tasks. Prompt+PTM tuning differs from the ``pre-train, fine-tune the holistic model" strategy by providing additional prompts that can provide additional bootstrapping at the start of model training. For example, \citet{TOIS23-PEPLER} proposed a continuous prompt learning approach by first fixing the PTM, tuning the prompt to bridge the gap between the continuous prompts and the loaded PTM, and then fine-tuning both the prompt and PTM, resulting in a higher BLUE score in empirical results. They combined both discrete prompts (three user/item feature keywords, such as gym, breakfast, and Wi-Fi) and soft prompts (user/item embeddings) to generate recommendation explanations. Case studies showed improvements in the readability and fluency of generated explanations using the proposed prompts. Note that the Prompt+PTM tuning stage does not necessarily mean the fine-tuning stage but can be any possible stage for tuning parameters from both sides for specific data input. \citet{xin2022rethinking} adapted a reinforcement learning framework as a Prompt+PTM tuning strategy by learning reward-state pairs as soft prompt encodings w.r.t. observed actions during training. At the inference stage, the trained prompt generator can directly generate soft prompt embeddings for the recommendation model to generate actions (items).

\section{Learning Objectives of LMRS}
This section will overview several typical learning tasks and objectives of language models and their adaptations for different recommendation tasks. 

\subsection{Language modelling objectives to recommendation}
The expensive manual efforts required for annotated datasets have led many language learning objectives to adopt self-supervised labels, converting them to classic probabilistic density estimation problems. Among language modelling objectives, autoregressive, reconstruction, and auxiliary are three categories commonly used \citep{liu2023pre}. Here, we only introduce several language modelling objectives used for RSs.

\noindent\textbf{Partial/ Auto-regressive Modelling (P/AM)} Given a text sequence $\mathbf{X}_{1:T}=[x_1, x_2,\cdots x_T]$, the training objective of AM can be summarized as a joint negative log-likelihood of each variable given all previous variables:
\begin{small}
\begin{equation}
\mathcal{L}_{AM} = -\, \sum^{T}_{t=1}log\, p(x_t|\mathbf{X}_{<t-1})
\end{equation}
\end{small}

\noindent Modern LMRS typically utilize popular pre-trained left-to-right LMs such as GPT-2 \citep{hada2021rexplug} and DialoGPT \citep{wang2022recindial,wang2022towards} as the backbone for explainable and conversational recommendations, respectively, to avoid the laborious task of pre-training from scratch. While auto-regressive objectives can effectively model context dependency, the modelling context can only be accessed from one direction, primarily left-to-right. To address this limitation, PAM is introduced, which extends AM by enabling the factorization step to be a span. For each input $\mathbf{X}$, one factorization order $M$ is sampled. One popular PTM that includes PAM as an objective is UniLMv2 \citep{bao2020unilmv2}. The pre-trained UniLMv2 model can be utilized to initialize the news embedding model for news recommendation \citep{yu2021tiny}. 

Besides directly leveraging PTMs trained on textual inputs, some researchers apply this objective to train inputs with sequential patterns, such as graphs \citep{geng2022path} and user-item interactions \citep{zheng2022spatial}. These patterns serve as either scoring functions to select suitable paths from the start node/user to the end node/item or detectors to explore novel user-item pairs.

\begin{table*}[!ht]
\tiny
    \centering
    \setlength\extrarowheight{-0.6pt}
    \begin{tabular}{m{1.5cm}<{\centering}|m{1.9cm}|m{4.0cm}|m{2.0cm}|m{2.3cm}<{\centering}|m{1.6cm}<{\centering}}
        \hlineB{2}
        Training Strategy  &  \multicolumn{1}{c|}{Paper}  & \multicolumn{1}{c|}{Learning Objective} &  \multicolumn{1}{c|}{Recommendation Task} & Data Type &  \multicolumn{1}{c}{Source Code} \\
        \hlineB{1}
        \rowcolor{lightgray!50} \multicolumn{6}{c}{Pre-training \& Fine-tuning}\\ \hlineB{1}
        \multirow{5}{*}{\tabincell{c}{Pre-training w/o \\Fine-tuning}} & \cite{sun2019bert4rec} & Pre-train: MLM & Sequential RS & Sequential data &  \href{https://github.com/FeiSun/BERT4Rec}{Link} \\ 
        \cline{2-6} & \cite{geng2022path} & Pre-train: AM & Explainable RS & Graph &  N/A \\ 
        \cline{2-6} & \cite{de2021transformers4rec} & \tabincell{l}{Pre-train: AM + MLM + PerLM + RTD} & Session-based RS & Textual + Sequential data &  \href{https://github.com/NVIDIA-Merlin/Transformers4Rec/}{Link} \\  \hlineB{1}
        \multirow{10}{*}{\tabincell{c}{Fine-tuning \\Holistic Model}} & \cite{kang2021apirecx} & \tabincell{l}{Pre-train: cross-entropy\\Fine-tune: cross-entropy} & Cross-library API RS & Textual data (code) & \href{https://github.com/yuningkang/APIRecX}{Link}   \\ 
        \cline{2-6} & \cite{wang2022recindial} & \tabincell{l}{Pre-train: AM\\Fine-tune: AM + cross-entropy}  & Conversational RS & Textual data + Graph & \href{https://github.com/Lingzhi-WANG/PLM-BasedCRS}{Link}   \\ 
        \cline{2-6} & \cite{xiao2022training} & \tabincell{l}{Pre-train: AM + MLM\\Fine-tune: AM} & News RS & Textual + Sequential data &  \href{https://github.com/Microsoft/SpeedyRec}{Link}  \\ 
         \cline{2-6} & \cite{zhang2022twhin} & \tabincell{l}{Pre-train: MLM + NT-Xent\\Fine-tune: Negative Sampling Loss} & Social RS & Textual data &  \href{https://github.com/xinyangz/TwHIN-BERT}{Link}  \\
         \cline{2-6} & \cite{wang2023curriculum} & \tabincell{l}{Pre-train: MNP + MEP + cross-entropy +\\ Contrastive Loss; Fine-tune: cross-entropy} & Top-N RS & Graph &  N/A  \\ \hlineB{1}        
         \multirow{6}{*}{\tabincell{c}{Fine-tuning \\Partial Model}} & \cite{hou2022towards} & \tabincell{l}{Pre-train: Contrastive Loss\\Fine-tune: cross-entropy} & \tabincell{l}{Cross-domain RS\\Sequential RS} & Textual + Sequential data & \href{https://github.com/RUCAIBox/UniSRec}{Link} \\ 
        \cline{2-6} & \cite{yu2021tiny} & \tabincell{l}{Pre-train: MLM + AM\\Fine-tune: cross-entropy + MSE + InfoNCE} & News RS & Textual + Sequential data &  \href{https://github.com/yflyl613/Tiny-NewsRec}{Link}  \\ 
        \cline{2-6} & \cite{wu2022mm} & \tabincell{l}{Pre-train: MMM + MAP\\Fine-tune: cross-entropy} & News RS &  Sequential + Multi-modal data & \href{https://github.com/zcfinal/MM-Rec}{Link}  \\  \hlineB{1}
          \multirow{8}{*}{\tabincell{c}{Fine-tuning \\External Part}} & \cite{zhou2020s3} & \tabincell{l}{Pre-train: MIM\\Fine-tune: Pairwise Ranking Loss} & \tabincell{l}{Sequential RS} & Textual + Sequential data & \href{https://github.com/RUCAIBox/CIKM2020-S3Rec}{Link}  \\
        \cline{2-6} & \cite{liu2022boosting} & \tabincell{l}{Pre-train: MTP + cross-entropy\\Fine-tune: cross-entropy} & News RS & Textual + Sequential data &  \href{https://github.com/Jyonn/PREC}{Link}  \\ 
        \cline{2-6} & \cite{ijcai2019p825} & \tabincell{l}{Pre-train: binary cross-entropy\\Fine-tune: cross-entropy} & Medication RS &  Graph &  \href{https://github.com/jshang123/G-Bert}{Link}  \\ 
         \cline{2-6} & \cite{liu2022graph} & \tabincell{l}{Pre-train: binary cross-entropy\\Fine-tune: BPR + binary cross-entropy} & Top-N RS &Textual data + Graph &  \href{https://github.com/pretrain/pretrain}{Link}  \\ \hlineB{1}
         \rowcolor{lightgray!50} \multicolumn{6}{c}{Prompting}\\ \hlineB{1}
          \multirow{5}{*}{\tabincell{c}{Fixed-PTM \\Prompt Tuning}} & \cite{wang2022towards} & \tabincell{l}{Pre-train: AM + MLM + cross-entropy\\Prompt-tuning: AM + cross-entropy} & Conversational RS & Textual data &  \href{https://github.com/RUCAIBox/UniCRS}{Link}  \\ 
        \cline{2-6} & \cite{wu2022personalized} & \tabincell{l}{Pre-train: Pairwise Ranking Loss\\Prompt-tuning: Pairwise Ranking Loss + \\Contrastive Loss}  & \tabincell{l}{Cross-domain RS\\Sequential RS} & Textual + Sequential data & N/A \\  \hlineB{1}
         \multirow{2.5}{*}{\tabincell{c}{Fixed-prompt \\PTM Tuning}} & \cite{yang2022improving} & \tabincell{l}{Pre-train: AM + MLM\\PTM Fine-tune: AM + cross-entropy} & Conversational RS & Textual data &  \href{https://github.com/by2299/MESE}{Link}  \\ 
         \cline{2-6} & \cite{deng2022unified} & \tabincell{l}{Pre-train: AM; PTM Fine-tune: AM} & Conversational RS &  Textual data &  \href{https://github.com/dengyang17/UniMIND}{Link}  \\   \hlineB{1}
         \multirow{3}{*}{\tabincell{c}{Tuning-free \\Prompting}} & \cite{sileo2022zero} & Pre-train: AM & Zero-Shot RS & Textual data &  \href{https://colab.research.google.com/drive/1f1mlZ-FGaLGdo5rPzxf3vemKllbh2esT?usp=sharing\#scrollTo=20YBuD7jJNVI}{Link}  \\ 
        \cline{2-6} & \cite{geng2022recommendation} &Pre-train: AM &  \tabincell{l}{Zero-Shot  RS\\Cross-domain RS} & Textual + Sequential data & \href{https://github.com/jeykigung/P5}{Link}   \\
\hlineB{1}
       \multirow{3}{*}{\tabincell{c}{Prompt+PTM \\Tuning}} & \cite{TOIS23-PEPLER} &  \tabincell{l}{Pre-train: AM; Prompt-tuning: NLL \\Prompt+PTM tuning: NLL + MSE\\} & Explainable RS & Textual data &  \href{https://github.com/lileipisces/PEPLER}{Link}  \\ 
        \cline{2-6} & \cite{xin2022rethinking} & \tabincell{l}{Prompt+PTM tuning: cross-entropy} & Next Item RS &  Sequential data &  N/A \\  
        \hlineB{2}
    \end{tabular}
   \begin{tablenotes}
      \scriptsize
      \item Note: NT-Xent: Normalized Temperature-scaled Cross Entropy Loss; MMM: Masked Multi-modal Modelling; MAP: Multi-modal Alignment Prediction; MIM: Mutual Information Maximization Loss; MTP: Masked News/User Token Prediction; NLL: Negative Log-likelihood Loss.
    \end{tablenotes}
    \caption{A list of representative LMRS methods with open-source code.}
    \label{tab:booktabs}
    \vspace{-0.3cm}
\end{table*}

\begin{table*}[!ht]
\tiny
    \centering
    \setlength\extrarowheight{-0.4pt}
    \begin{tabular}{m{1.38cm}<{\centering}|m{1.6cm}<{\centering}|m{1.6cm}|m{5.2cm}|m{3.71cm}}
        \hlineB{2}
        Dataset  &  \multicolumn{1}{c|}{Data Source}  & \multicolumn{1}{c|}{Recommendation Task} &  \multicolumn{1}{c|}{Training Strategy} & Data Type \\
        \hlineB{1}
        \multirow{11}{*}{\tabincell{c}{MovieLens}} & \multirow{11}{*}{\tabincell{c}{\href{https://grouplens.org/datasets/movielens/25m/}{Link}}} & Rating Prediction & Tuning-free Prompting \cite{gao2023chat} & \multirow{11}{=}{Textual data \citep{zhang2021language,sileo2022zero,penha2020does,xie2022factual,gao2023chat}; \\Sequential data \cite{yuan2020future,liu2021pre,zhao2022resetbert4rec}; \\Graph \cite{liu2022graph,liu2021pre,wang2023curriculum}; \\Multi-modal data \cite{liu2021pre}} \\ 
        \cline{3-4} & & Explainable RS & Fine-tuning Holistic Model \cite{xie2022factual} &  \\ 
        \cline{3-4} & & Sequential RS & Pre-training w/o Fine-tuning \cite{yuan2020future}, Fine-tuning Holistic Model \cite{zhao2022resetbert4rec} &  \\  
        \cline{3-4} & & Conversational RS & Fine-tuning Holistic Model \cite{penha2020does}, Tuning-free Prompting \cite{gao2023chat} & \\
        \cline{3-4} & & Top-N RS & Fine-tuning Holistic Model \cite{wang2023curriculum}, Fine-tuning External Part \cite{liu2022graph}, Fixed-prompt PTM Tuning \citep{zhang2021language}, Tuning-free Prompting \citep{zhang2021language,sileo2022zero} &  \\
        \cline{3-4} & & CTR Prediction & Fine-tuning External Part \cite{liu2021pre} &  \\ \hlineB{1} 
        \multirow{17}{*}{\tabincell{c}{Amazon \\Review Data}} & \multirow{17}{*}{\tabincell{c}{\href{http://jmcauley.ucsd.edu/data/amazon/}{Link}}} & Rating Prediction & Fine-tuning External Part \cite{hada2021rexplug}, Tuning-free Prompting \cite{geng2022recommendation} & \multirow{17}{=}{Textual data \cite{hada2021rexplug,qiu2021u,TOIS23-PEPLER,geng2022recommendation,zhou2020s3,penha2020does,xie2022factual,zhao2022resetbert4rec,Hou2022LearningVI,Li2023GPT4RecAG}; \\Sequential data \cite{sun2019bert4rec,geng2022recommendation,zhou2020s3,geng2022path,liu2021pre,Hou2022LearningVI,Guo2023AutomatedPF}; \\Graph \cite{geng2022path,liu2021pre}; \\Multi-modal data \cite{liu2021pre}}  \\ 
        \cline{3-4} &  & Cross-domain RS  & Fine-tuning Holistic Model \cite{qiu2021u}, Fine-tuning Partial Model \cite{Hou2022LearningVI}, Fixed-PTM Prompt Tuning \cite{Guo2023AutomatedPF} &   \\ 
        \cline{3-4} & & Explainable RS & Pre-training w/o Fine-tuning \cite{geng2022path}, Fine-tuning Holistic Model \cite{xie2022factual}, Fixed-PTM Prompt Tuning \cite{TOIS23-PEPLER}, Fixed-prompt PTM Tuning \cite{Li2023GPT4RecAG}, Tuning-free Prompting \cite{geng2022recommendation}&   \\ 
        \cline{3-4} &  & Zero-Shot RS  & Tuning-free Prompting \cite{geng2022recommendation} &   \\ 
        \cline{3-4} &  & Sequential RS  & Pre-training w/o Fine-tuning \cite{sun2019bert4rec}, Fine-tuning Holistic Model \cite{zhao2022resetbert4rec}, Fine-tuning Partial Model \cite{Hou2022LearningVI}, Fine-tuning External Part \cite{zhou2020s3}, Fixed-PTM Prompt Tuning \cite{Guo2023AutomatedPF}, Tuning-free Prompting \cite{geng2022recommendation} &   \\ 
        \cline{3-4} &  & Conversational RS  & Fine-tuning Holistic Model \cite{penha2020does} &   \\
        \cline{3-4} &  & Top-N RS  & Fine-tuning External Part \cite{liu2021pre} &   \\
        \hlineB{1}
        \multirow{14}{*}{\tabincell{c}{Yelp}} & \multirow{14}{*}{\tabincell{c}{\href{https://www.kaggle.com/datasets/yelp-dataset/yelp-dataset}{Link}}} & Rating Prediction & Fine-tuning Holistic Model \cite{xie2022factual}, Fine-tuning External Part \cite{hada2021rexplug,geng2022improving}, Tuning-free Prompting \cite{geng2022recommendation}& \multirow{14}{=}{Textual data \cite{hada2021rexplug,qiu2021u,TOIS23-PEPLER,geng2022recommendation,xiao2021uprec,zhou2020s3,xie2022factual};\\Sequential data \cite{geng2022recommendation,xiao2021uprec,zhou2020s3,sankar2021protocf};\\Graph \cite{xiao2021uprec,zheng2022spatial,wang2023curriculum}; \\Multi-modal data \cite{geng2022improving}}  \\
        \cline{3-4} &  & Cross-domain RS  & Fine-tuning Holistic Model \cite{qiu2021u} &   \\ 
        \cline{3-4} & & Explainable RS & Fine-tuning Holistic Model \cite{xie2022factual}, Fine-tuning
External Part \cite{geng2022improving}, Fixed-PTM Prompt Tuning \cite{TOIS23-PEPLER}, Tuning-free Prompting \cite{geng2022recommendation} &   \\ 
        \cline{3-4} &  & Zero-Shot RS  & Tuning-free Prompting \cite{geng2022recommendation} &   \\ 
        \cline{3-4} &  & Sequential RS  & Fine-tuning Holistic Model \cite{xiao2021uprec}, Fine-tuning External Part \cite{zhou2020s3}, Tuning-free Prompting \cite{geng2022recommendation} &   \\ 
        \cline{3-4} &  & Top-N RS  & Pre-training w/o Fine-tuning \cite{zheng2022spatial}, Fine-tuning Holistic Model \cite{wang2023curriculum}, Fine-tuning External Part \cite{sankar2021protocf} &   \\
        \hlineB{1}
        \multirow{4}{*}{\tabincell{c}{TripAdvisor}} & \multirow{4}{*}{\tabincell{c}{\href{https://lifehkbueduhk-my.sharepoint.com/:f:/g/personal/16484134_life_hkbu_edu_hk/Eln600lqZdVBslRwNcAJL5cBarq6Mt8WzDKpkq1YCqQjfQ?e=cISb1C}{Link}}} & Rating Prediction & Fine-tuning Holistic Model \cite{xie2022factual}, Fine-tuning External Part \cite{geng2022improving} & \multirow{4}{=}{Textual data \cite{TOIS23-PEPLER,xie2022factual}; \\Multi-modal data \cite{geng2022improving}}  \\
        \cline{3-4} &  & Explainable RS  & Fine-tuning Holistic Model \cite{xie2022factual}, Fine-tuning External Part \cite{geng2022improving}, Fixed-PTM Prompt Tuning \cite{TOIS23-PEPLER} &   \\
        \hlineB{1}
        MIND & \tabincell{c}{\href{https://msnews.github.io/}{Link}} & Top-N RS & Fine-tuning Holistic Model \cite{xiao2022training}, Fine-tuning Partial Mode \cite{yu2021tiny}, Fine-tuning External Part \cite{yu2021tiny}, Fixed-prompt PTM Tuning \cite{Zhang2023PromptLF} & Textual data \cite{xiao2022training,yu2021tiny,Zhang2023PromptLF}; Sequential data \cite{xiao2022training,yu2021tiny}  \\
        \hlineB{1}
        ReDial & \tabincell{c}{\href{https://redialdata.github.io/website/download}{Link}} & Conversational RS & Fine-tuning Holistic Model \cite{li2022self}, Fixed-PTM Prompt Tuning \cite{wang2022towards}, Fixed-prompt PTM Tuning \cite{yang2022improving} & Textual data \cite{wang2022towards,yang2022improving,li2022self}; Graph \cite{li2022self}  \\
        \hlineB{1}
        Polyvore Outfits & \tabincell{c}{\href{https://github.com/xthan/polyvore/tree/master/data}{Link}} & Fashion RS & Fine-tuning Partial Model + External Part \cite{sarkar2022outfittransformer} & Multi-modal data \cite{sarkar2022outfittransformer}  \\
        \hlineB{1}
        MIMIC-III & \tabincell{c}{\href{https://www.nature.com/articles/sdata201635\#MOESM102}{Link}} & Medication RS & Fine-tuning External Part \cite{ijcai2019p825} & Graph \cite{ijcai2019p825}  \\
        \hlineB{1}
        Stackoverflow & \tabincell{c}{\href{https://archive.org/details/stackexchange}{Link}} & Top-N RS & Fine-tuning Holistic Mode \cite{he2022ptm4tag} & Textual data \cite{he2022ptm4tag}  \\
        \hlineB{1}
        Online Retail & \tabincell{c}{\href{https://www.kaggle.com/datasets/carrie1/ecommerce-data}{Link}} & Cross-domain RS & Fine-tuning Partial Model \cite{hou2022towards} & Textual + Sequential data \cite{hou2022towards}  \\
        \hlineB{2}
    \end{tabular}
    \caption{A list of commonly used and publicly accessible real-world datasets for LMRS.}
    \label{tab:dataset}
    \vspace{-0.3cm}
\end{table*}

\noindent\textbf{Masked Language Modelling (MLM)} 
Taking a sequence of textual sentences as input, MLM first masks a token or multi-tokens with a special token such as $[MASK]$. Then the model is trained to predict the masked tokens taking the rest of the tokens as context. The objective is as follows:
\begin{small}
\begin{equation}
\mathcal{L}_{MLM} = -\, \sum_{\hat{x}\in M(\mathbf{X})}log\, p(\hat{x}|\mathbf{X}_{\\M(\mathbf{X})})
\end{equation}
\end{small}
\noindent where $M(\mathbf{X})$ and $\mathbf{X}_{\\M(\mathbf{X})}$ represent the masked tokens in the input sequence $\mathbf{X}$ and the rest of the tokens in $\mathbf{X}$ respectively. A typical example of MLM training strategy can be found on BERT, which is leveraged as backbone in \citep{ijcai2021p462} to capture user-news matching signals for news recommendation.

Concurrently, some research works propose multiple enhanced versions of MLM. RoBERTa \citep{liu2019roberta} improves BERT by dynamic masking instead of static manner and can be used to initiate word embedding for conversations \citep{wang2022towards} and news articles \citep{wu2021empowering} for different recommendation scenarios.

\noindent\textbf{Next Sentence Prediction (NSP)} It is a binary classification loss for predicting whether two segments follow each other in the original text. The training can be performed in a self-supervised way by taking positive examples from consecutive sentences from the input text corpus and creating negative examples by pairing segments from different documents. A general loss of the NSP is as follows:
\begin{small}
\begin{equation}
\mathcal{L}_{NSP} = -\, log\, p(c|\mathbf{x},\mathbf{y})
\end{equation}
\end{small}
\noindent where $\mathbf{x}$ and $\mathbf{y}$ represent two segments from the input corpus, and $c=1$ if $\mathbf{x}$ and $\mathbf{y}$ are consecutive, otherwise $c=0$. The NSP objective involves reasoning about the relationships between pairs of sentences and can be utilized for better representation learning of textual items such as news articles, item descriptions, and conversational data for recommendation purposes. Moreover, it can be employed to model the intimate relationships between two components. \citet{malkiel2020recobert} used the NSP to capture the relationship between the title and description of an item for next-item prediction. Furthermore, models pre-trained with NSP (such as BERT) can be leveraged for probing the learned knowledge with prompts, which are then infused in the fine-tuning stage to improve model training on adversarial data for conversational recommendation \citep{penha2020does}. Sentence Order Prediction (SOP) as a variation of the NSP takes two consecutive segments from the same document as positive examples, which are then swapped in order as negative examples. SOP has been used to learn the inner coherence of title, description, and code for tag recommendation on StackOverflow \citep{he2022ptm4tag}.

Nevertheless, some researchers have questioned the necessity and effectiveness of the NSP and SOP for downstream tasks \citep{he2022ptm4tag}, which highlights the need for further investigation in recommendation scenarios.

\noindent\textbf{Replaced Token Detection(RTD)} It is used to predict whether a token is replaced given its surrounding context:
\begin{small}
\begin{equation}
\mathcal{L}_{RTD} = -\, \sum^{T}_{t=1}log\, p(y_t|\mathbf{\hat{X}})
\end{equation}
\end{small}
where $y_t=\mathbf{1} (\hat{x}_t=x_t)$, and $\mathbf{\hat{X}}$ is corrupted from the input sequence $\mathbf{X}$. \citet{de2021transformers4rec} trained a Transformer-based model with RTD objective for session-based recommendations, which achieved the best performance among MLM and AM objectives. This is probably because RTD takes the whole user-item interaction sequence as input and model the context from the bidirectional way.

\subsection{Adaptive objectives to recommendation}
Numerous pre-training or fine-tuning objectives draw inspiration from LM objectives and have been effectively applied to specific downstream tasks based on the input data types and recommendation goals. In sequential recommendations, there is a common interest in modelling an ordered input sequence in an auto-regressive manner from left to right.

Analogous to text sentences, \citet{zheng2022spatial} and \citet{xiao2022training} treated the user's clicked news history as input text and proposed to model user behavior in an auto-regressive manner for next-click prediction. However, as the sequential dependency may not always hold strictly in terms of user preference for recommendations \citep{yuan2020future}, MLM objectives can be modified accordingly. \citet{yuan2020parameter} randomly masked a certain percentage of historical user records and predicted the masked items during training. Auto-regressive learning tasks can also be adapted to other types of data. \citet{geng2022path} modeled a series of paths sampled from a knowledge graph in an auto-regressive manner for recommendation by generating the end node from the pre-trained model. \citet{zhao2022resetbert4rec} proposed pre-training the Rearrange Sequence Prediction task to learn the sequence-level information of the user's entire interaction history by predicting whether the user interaction history had been rearranged, which is similar to Permuted Language Modelling (PerLM) \citep{yang2019xlnet}.

MLM, also known as Cloze Prediction, can be adapted to learn graph representations for different recommendation purposes. \citet{wang2023curriculum} proposed pre-training a transformer model on a reconstructed subgraph from a user-item-attribute heterogeneous graph, using Masked Node Prediction (MNP), Masked Edge Prediction (MEP), and meta-path type prediction as objectives. Specifically, MNP was performed by randomly masking a proportion of nodes in a heterogeneous subgraph and then predicting the masked nodes based on the remaining contexts by maximizing the distance between the masked node and the irrelevant node. Similarly, MEP was used to recover the masked edge of two adjacent nodes based on the surrounding context. Apart from that, MLM can also be adapted to multi-modal data called Masked Multi-modal Modelling (MMM) \citep{wu2022mm}. MMM was performed by predicting the semantics of masked news and news image regions given the unmasked inputs and indicating whether a news image and news content segment correspond to each other for news recommendation purposes.

The NSP/SOP can be adapted for CTR prediction as Next K Behaviors Prediction (NBP). NBP was proposed to learn user representations in the pre-training stage by inferring whether a candidate behavior is the next $i$-th behavior of the target user based on their past N behaviors. NBP can also capture the relatedness between past and multiple future behaviors.

\section{Formulating Training with Data Types}
To associate training strategy, learning objectives with different input data types, we summarize representative works in this domain in Table \ref{tab:booktabs}. The listed training strategies and objectives are carefully selected and are typical in existing work. For the page limit, we only selected part of recent research on LMRS. For more research progress and related resources, please refer to \url{https://github.com/SmartmediaAI/LMRS}. 

Considering that datasets are another important factor for empirical analysis of LMRS approaches, in Table \ref{tab:dataset}, we also list several representative publicly available datasets taking into account the popularity of data usage and the diversity of data types, as well as their corresponding recommendation tasks, training strategies, and adopted data types. From Table \ref{tab:dataset}, We draw several observations: First, datasets can be converted into different data types, which can then be analyzed from various perspectives to enhance downstream recommendations. The integration of different data types can also serve different recommendation goals more effectively \citep{geng2022recommendation,liu2021pre}. For instance, \citet{liu2021pre} transformed user-item interactions and multimodal item side information into a homogeneous item graph. A sampling approach was introduced to select and prioritize neighboring nodes around a central node. This process effectively translated the graph data structure into a sequential format. The subsequent training employed a self-supervised signal within a transformer framework, utilizing an objective for reconstructing masked node features. The resultant pre-trained node embeddings could be readily applied for recommendation purposes, or alternatively, fine-tuned to cater to specific downstream objectives. Second, some training strategies can be applied to multiple downstream tasks by fine-tuning a few parameters from the pre-trained model, adding an extra component, or using different prompts. \citet{geng2022recommendation} designed different prompt templates for five different tasks to train a transformer-based model with a single objective, and achieved improvements on multiple tasks with zero-shot prompting. \citet{deng2022unified} unified the multiple goals of conversational recommenders into a single sequence-to-sequence task with textual input, and designed various prompts to shift among different tasks. We further observe that prompting methods are primarily used in LMRS with textual and sequential data types, but there has been a lack of exploration for multi-modal or graph data. This suggests that investigating additional data types may be a future direction for research in prompting-based LMRS.

\section{Evaluation}
\begin{figure}
\centering
   \epsfig{figure=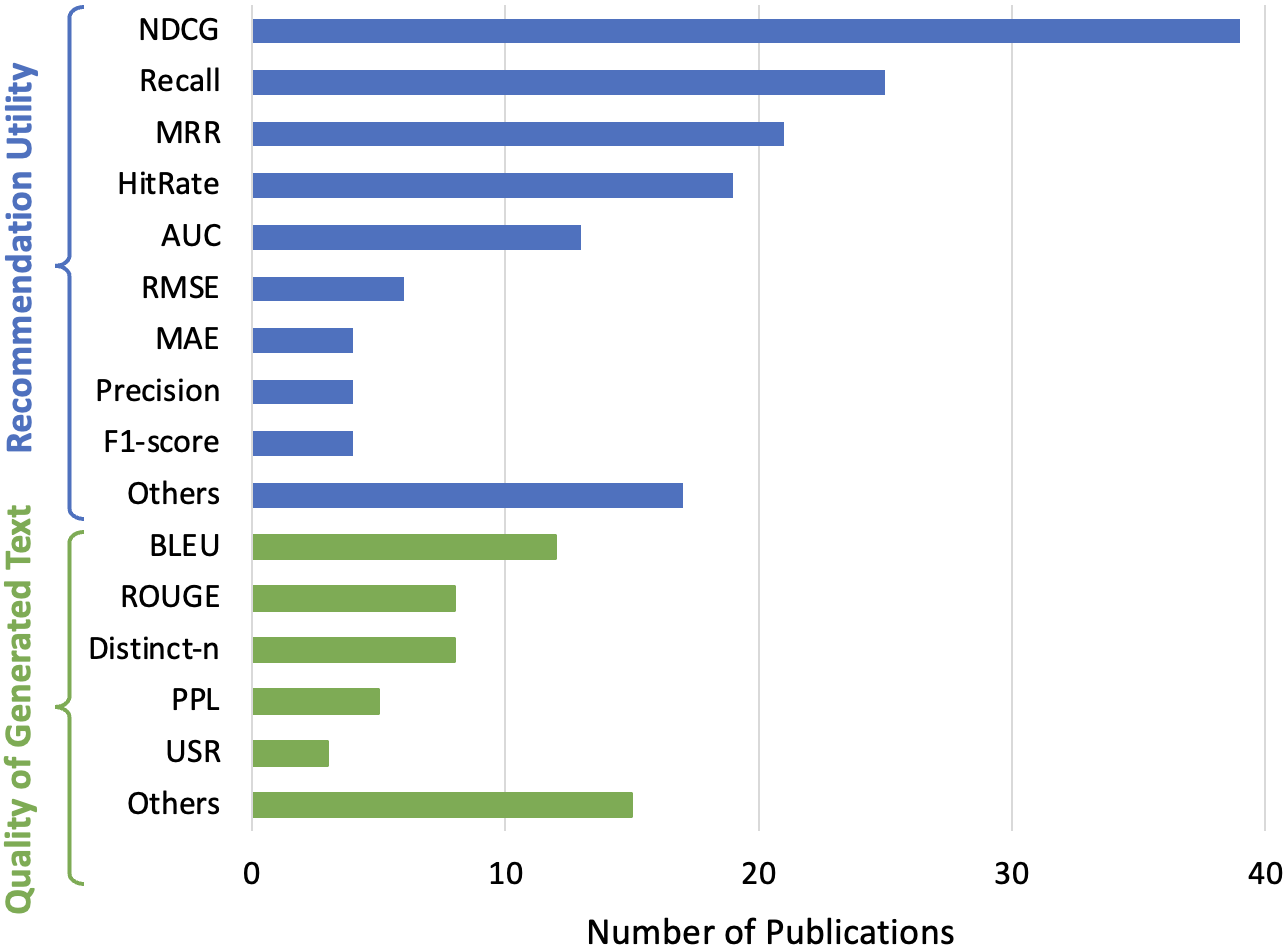,height=132pt,width=189pt}
\caption{The statistics of evaluation metrics on recommendation utility and generated text quality in LMRS.} \label{fig:evaluation}
\end{figure}
\subsection{Evaluation metrics}
As an essential aspect of recommendation design, evaluation can provide insights on recommendation quality from multiple dimensions. Apart from well-known metrics such as RMSE, MAP, AUC, MAE, Recall, Precision, MRR, NDCG, F1-score and HitRate in offline mode, some works define Group AUC \citep{zhang2022keep} or User Group AUC \citep{zheng2022spatial} to evaluate the utility of group recommendations. \citet{jiang2022learning} and \citet{liu2022boosting} conducted A/B testing to evaluate performance with online users using Conversion rate or CTR. 

The integration of generative modules such as GPT and T5 into existing recommender systems offers additional possibilities for recommender systems, such as generating free-form textual explanations for recommendation results or simulating more realistic real-life dialogue scenarios during conversational recommendations to enhance users' experience. In such cases, BLEU and ROUGE are commonly adopted to automatically evaluate the relevance of generated text based on lexicon overlap. Besides, Perplexity (PPL), Distinct-n, and Unique Sentence Ratio (USR) are also widely used metrics to measure fluency, diversity, and informativeness of generated texts. Other evaluation metrics are leveraged with respect to special requests in LMRSs. For instance, \citet{xie2022factual} adopted Entailment Ratio and MAUVE to measure if the generated explanations are factually correct and how close the generated contents are to the ground truth corpus, respectively. \citet{geng2022improving} adopted Feature Diversity (DIV) and CLIPScore (CS) to measure the generated explanations and text-image alignment. Besides, to assess the system's capability to provide item recommendations during conversations, \citet{wang2022recindial} computed the Item Ratio within the final generated responses. They evaluated the recommendation performance in an end-to-end manner to prevent the inappropriate insertion of recommended items into dialogues.

Human evaluation complements objective evaluation, as automatic metrics may not match subjective feedback from users. \citet{liu2023chatgpt} pointed out that human subjective and automatic objective evaluation measurements may yield opposite results, which underscores the limitations of existing automatic metrics for evaluating generated explanations and dialogues in LMRSs. Figure \ref{fig:evaluation} displays usage frequency statistics for different evaluation metrics in their respective tasks.

\begin{table*}[!ht]
\tiny
    \centering
    \begin{tabular}{m{0.9cm}|m{0.6cm}|m{0.6cm}|m{0.6cm}|m{0.6cm}|m{0.6cm}|m{0.6cm}|m{0.6cm}|m{0.6cm}|m{0.6cm}|m{0.6cm}|m{0.6cm}|m{0.6cm}}
        \hlineB{2}
        \multirow{3}{*}{Metrics}  &  \multicolumn{6}{c|}{Fine-tune Holistic Model}  & \multicolumn{3}{c|}{Fixed-PTM Prompt Tuning} &  \multicolumn{3}{c}{Fixed-prompt PTM Tuning}  \\      
         \cline{2-13} &  \multicolumn{3}{c|}{\citep{yang2022improving}}  & \multicolumn{3}{c|}{\citep{li2022self}} &  \multicolumn{3}{c|}{\citep{wang2022towards}} & \multicolumn{3}{c}{\citep{yang2022improving}} \\ 
        \cline{2-13} & ReDial & KBRD & KGSF & ReDial & KBRD & KGSF & ReDial & KBRD & KGSF & ReDial & KBRD & KGSF \\ \hlineB{1}
        Recall@1 & \textbf{1.458} & \textbf{0.903} & \textbf{0.513} & -- & -- & -- & 1.217 & 0.545 & 0.457 & 1.333 & 0.860 & 0.436 \\ 
        Recall@10 & 0.174 & 0.6 & 0.311 & 0.307 & 0.219 & 0.115 & 0.736 & 0.28 & 0.266 & \textbf{0.829} & \textbf{0.707} & \textbf{0.399} \\ 
        Recall@50 & 0.291 & 0.229 & 0.093 & 0.268 & 0.154 & 0.043 & \textbf{0.439} & 0.248 & 0.128 & 0.422 & \textbf{0.354} & \textbf{0.204} \\ \hlineB{1}
        Distinct-2 & 1.031 & 0.738 & 0.581 & 0.541 & 0.149 & 0.159 & 1.187 & 0.751 & 0.629 & \textbf{2.653} & \textbf{2.125} & \textbf{1.844}\\ 
        Distinct-3 & 1.767 & 0.774 & 0.505 & 1.408 & 0.492 & 0.204 & 1.746 & 0.71 & 0.497 & \textbf{3.881} & \textbf{2.13} & \textbf{1.654} \\ 
        Distinct-4 & 2.338 & 0.799 & 0.466 & 1.524 & 0.7 & 0.225 & 2.649 & 0.9 & 0.597 & \textbf{4.759} & \textbf{2.104} & \textbf{1.530} \\ 
        \hlineB{2}
    \end{tabular}
    \caption{LMRSs performance comparison using common benchmarks on the ReDial dataset.}
    \label{tab:comparison1}
\end{table*}

\begin{table*}[!ht]
\tiny
    \centering
    \begin{tabular}{m{1.0cm}|m{1.1cm}|m{0.7cm}|m{0.5cm}|m{0.7cm}|m{0.7cm}|m{0.5cm}|m{0.5cm}|m{0.5cm}|m{0.5cm}|m{0.7cm}|m{0.7cm}|m{0.7cm}|m{0.7cm}}
        \hlineB{2}
       \multirow{2}{=}{Training Strategy} & \multirow{2}{*}{Paper} & \multicolumn{4}{c|}{Caser}  & \multicolumn{4}{c|}{GRU4Rec}  &  \multicolumn{4}{c}{SASRec}  \\      
        \cline{3-14} & & H@5 & N@5 &  H@10 & N@10 & H@5 & N@5 &  H@10 & N@10 & H@5 & N@5 &  H@10 & N@10 \\ \hlineB{1}
      Pre-train & \citep{sun2019bert4rec} & 0.3582 & 0.5229 & 0.168 & 0.3691 & 0.2392 & 0.3643 & 0.1398 & 0.2815 & 0.1412 & 0.1135 & 0.1402 & 0.1402\\ \hlineB{1}
      Fine-tune Extra Part & \citep{zhou2020s3} & 0.4848 & 0.5354 & 0.3968 & 0.4857 & 0.4406 & 0.5022 & 0.341 & 0.4443 & 0.2034 & 0.1963 & \textbf{0.1725} & 0.1825\\  \hlineB{1}
       \multirow{3}{=}{Tuning-free Prompt} & \citep{geng2022recommendation} & \textbf{1.478} & \textbf{1.8931} & \textbf{0.9135} & \textbf{1.4375} & \textbf{2.0976} & \textbf{2.8283} & \textbf{1.3463} &\textbf{2.1314} & \textbf{0.3127} & \textbf{0.5221} & 0.0975 & \textbf{0.3491}\\ 
        \cline{2-14} & \citep{liu2023chatgpt} &-0.3415 & 0.0305 & -0.611 & -0.233 & -- & -- & -- & -- & -0.6512 & -0.4578 & -0.7769 & -0.5755\\
        \hlineB{2}
    \end{tabular}
    \caption{LMRSs performance comparison using common benchmarks on the Amazon Beauty dataset.}
    \label{tab:comparison2}
\end{table*}

\begin{table*}[!ht]
\tiny
    \centering
    \begin{tabular}{m{1.0cm}|m{1.1cm}|m{0.4cm}|m{0.4cm}|m{0.4cm}|m{0.45cm}|m{0.4cm}|m{0.4cm}|m{0.4cm}|m{0.45cm}|m{0.55cm}|m{0.55cm}|m{0.4cm}|m{0.45cm}|m{0.4cm}|m{0.4cm}|m{0.4cm}|m{0.45cm}}
        \hlineB{2}
       \multirow{2}{=}{Training Strategy} & \multirow{2}{*}{Paper} & \multicolumn{4}{c|}{Caser}  & \multicolumn{4}{c|}{SASRec}  &  \multicolumn{4}{c|}{BERT4Rec} & \multicolumn{4}{c}{GRU4Rec}  \\      
        \cline{3-18} & & H@5 & N@5 &  H@10 & N@10 & H@5 & N@5 &  H@10 & N@10 & H@5 & N@5 &  H@10 & N@10 & H@5 & N@5 &  H@10 & N@10 \\ \hlineB{1}
      Fine-tune Holistic Model & \citep{xiao2021uprec} & 0.2097 & 0.1953 & 0.2078 & 0.1966 & 0.2581 & 0.2380 & 0.2811 & 0.2533 & 0.0666 & 0.087 & 0.617 & 0.081 & 0.3022 & 0.3961 & 0.2.26 & 0.3153\\ \hlineB{1}
      Fine-tune Extra Part & \citep{zhou2020s3} & 0.1906 & 0.178 & 0.1597 & 0.1753 & 0.0592 & 0.07 & 0.0477 & 0.0629 & 0.0182 & 0.035 & 0.0168 & 0.0326 & 0.1192 & 0.1631 & 0.0633 & 0.1278\\  \hlineB{1}
      Tuning-free Prompt & \citep{geng2022recommendation} & \textbf{2.8013} & \textbf{3.1979} & \textbf{1.7945} & \textbf{2.4651} & \textbf{2.5215} & \textbf{3.03} & \textbf{1.5803} & \textbf{2.2868} & \textbf{10.2549} & \textbf{11.2121} & \textbf{6.8556} & \textbf{8.9333} & \textbf{2.7763} & \textbf{3.0707} & \textbf{1.6882} & \textbf{2.3358}\\
        \hlineB{2}
    \end{tabular}
    \caption{LMRSs performance comparison using common benchmarks on the Yelp dataset.}
    \label{tab:comparison3}
\end{table*}

\begin{table*}[!ht]
\tiny
    \centering
    \begin{tabular}{m{1.0cm}|m{1.1cm}|m{0.4cm}|m{0.4cm}|m{0.4cm}|m{0.45cm}|m{0.4cm}|m{0.4cm}|m{0.4cm}|m{0.45cm}|m{0.4cm}|m{0.4cm}|m{0.4cm}|m{0.45cm}|m{0.4cm}|m{0.4cm}|m{0.4cm}|m{0.45cm}}
        \hlineB{2}
       \multirow{2}{=}{Training Strategy} & \multirow{2}{*}{Paper} & \multicolumn{4}{c|}{NAML}  & \multicolumn{4}{c|}{NPA}  &  \multicolumn{4}{c|}{LSTUR} & \multicolumn{4}{c}{NRMS}  \\      
        \cline{3-18} & & AUC & MRR &  N@5 & N@10 & AUC & MRR & N@5 & N@10 & AUC & MRR & N@5 & N@10 & AUC & MRR & N@5 & N@10 \\ \hlineB{1}
       \multirow{2}{=}{Fine-tune Holistic Model} & \citep{ijcai2021p462} & 0.0635 & 0.0895 & 0.0973 & 0.0816 & 0.0722 & 0.1126 & 0.127 & 0.1092 & 0.0537 & 0.1026 & 0.1132 & 0.0941 & 0.0446 & 0.0731 & 0.0786 & 0.0667\\
       \cline{2-18} & \citep{xiao2022training} & \textbf{0.0913} & \textbf{0.1784} & \textbf{0.1974} & \textbf{0.1713} & \textbf{0.1343} & \textbf{0.2855} & \textbf{0.32} & \textbf{0.2793} & \textbf{0.1456} & \textbf{0.3018} & \textbf{0.3448} & \textbf{0.2906} & \textbf{0.0746} & \textbf{0.1612} & \textbf{0.1825} & \textbf{0.1575}\\  \hlineB{1}
       \multirow{2}{=}{Fine-tune Partial/ Extra Part} & \citep{wu2021empowering} & 0.0401 & 0.0608 & 0.0666 & 0.0553 & 0.039 & 0.063 & 0.0654 & 0.0538 & 0.037 & 0.0594 & 0.0659 & 0.0525 & 0.0361 & 0.0631 & 0.0661 & 0.0517\\
        \cline{2-18} & \citep{shin2023scaling} & -- & -- & -- & -- & 0.0772 & 0.1416 & 0.1557 & 0.1231 & 0.0572 & 0.1131 & 0.1281 & 0.1041 & 0.0611 & 0.1066 & 0.1222 & 0.094\\
        \hlineB{2}
    \end{tabular}
    \caption{LMRSs performance comparison using common benchmarks on the MIND dataset.}
    \label{tab:comparison4}
\end{table*}

\subsection{Discussion on evaluation across datasets}
In this section, we compare the results obtained from various models using commonly used datasets. Specifically, based on the reported results in the paper, we measured the improvement achieved by different models compared to a shared baseline and evaluated them using the same metrics on the same dataset. The comparisons are presented in Table \ref{tab:comparison1}$\sim$\ref{tab:comparison4}. Most improvements are highlighted in bold, and N@k denotes NDCG@k, H@k denotes HitRate@k. It's important to recognize that a comprehensive and precise assessment cannot be achieved without a carefully designed platform and thoughtful settings for conducting the experiments. Various factors, such as diverse training platforms, parameter settings, and data split strategy, can lead to fluctuations in the results. Hence, it is essential to consider the analysis solely for reference purposes. From the tables, we can observe that:  First, among the four conversational recommender systems assessed using the ReDial dataset, \textit{fixed prompt PTM tuning} paradigm \citet{yang2022improving} demonstrate the most significant improvements compared to the shared baselines. Second, on the Amazon dataset, \textit{zero-shot} and \textit{few-shot learning} of ChatGPT underperformed the supervised recommendation baselines \citep{liu2023chatgpt}. This could be due to language models' strength in capturing language patterns rather than effectively collaborating to suggest similar items based on user preferences \citep{zhang2021language}. Besides, \citet{liu2023chatgpt} pointed out that the position of candidate items in the item pool can also affect the direct recommendation performance. Another \textit{prompting-based} model, P5, showed the most improvements for both Amazon and Yelp datasets \citep{geng2022recommendation}, which verifies the need for more guidance when using large pre-trained language models for recommendations. Finally, for news recommendation on the MIND dataset, \citet{xiao2022training} introduced a model-agnostic \textit{fine-tuning} framework with cache management, which can accelerate the model training process and yield the most improvements over the baselines.

\section{Discussion and Future Directions}
Despite the effectiveness of LM training paradigms has been verified in various recommendation tasks, there are still several challenges that could be the future research directions.

\noindent\textbf{Language bias and fact-consistency in language generation tasks of recommendation.}
While generating free-form responses of conversational recommender systems or explanations of the recommended results, the generative components of existing LMRSs tend to predict generic tokens to ensure sentences fluency or repeat certain universally applicable ``safe" sentences (e.g. ``the hotel is very nice" generated from PETER \citep{li2021personalized}). Therefore, one future research direction is to enhance the diversity and pertinence of generated explanations and replies while maintaining language fluency, rather than resorting to "Tai Chi" responses. Additionally, generating factually consistent sentences is also an urgent research problem that needs to be addressed but has not received sufficient attention \citep{xie2022factual}.

\noindent\textbf{Knowledge transmission and injection for downstream recommendations.}
Improper training strategies may cause varying degrees of problems when transferring knowledge from pre-trained models. \citet{zhang2022keep} have pointed out the catastrophic forgetting problem in continuously-trained industrial recommender systems. The degree of domain knowledge pre-trained models possess and the effective ways to transfer and inject it for recommendation purposes are both open questions. For example, \citet{zhang2021language} experimented with a simple approach to injecting knowledge through domain-adaptive pre-training, resulting in only limited improvements. Furthermore, questions about maximizing knowledge transfer to different recommendation tasks, quantifying the degree of transferred knowledge, and whether an upper bound for knowledge transfer exists are all valuable issues that need to be studied and explored in the AI community.

\noindent\textbf{Scalability of pre-training mechanism in recommendation.}
As model parameters growing larger and larger, the knowledge stored in them is also increasing. Despite the great success of pre-trained models in multiple recommendation tasks, how to maintain and update such complex and large-scale models without affecting the efficiency and accuracy of recommendations in reality needs more attention. Some works have proposed improving model updating efficiency by fine-tuning a partial pre-trained model or an extra part with far fewer parameters than the model's magnitude. However, \citet{yuan2020parameter} empirically found that fine-tuning only the output layer often resulted in poor performance in recommendation scenarios. While properly fine-tuning the last few layers sometimes offered promising performance, the improvements were quite unstable and depended on the pre-trained model and tasks. \citet{yu2021tiny} proposed compressing large pre-trained language models into student models to improve recommendation efficiency, while \citet{yang-etal-2022-gram} focused on accelerating the fine-tuning of pre-trained language models and reducing GPU memory footprint for news recommendation by accumulating the gradients of redundant item encodings. Despite all these achievements, efforts are still needed in this rapidly developing field.

\noindent\textbf{Balancing multiple objectives in pre-training}
Many research works use multi-task learning objectives to better apply the knowledge learned in the pre-training phase to downstream tasks \citep{geng2022recommendation, wang2023curriculum}. The primary objective of multi-task learning for recommendation is to enhance recommendation accuracy and/or other related aspects by promoting interactions among related tasks. The learning optimization process requires trade-offs among different objectives. For instance, \citet{wang2023quotation} fine-tuned parameters to optimize and balance the overarching goals of topic-level recommendation, semantic-level recommendation, and a specific aspect of topic learning. Similarly in \citep{wang2022learning}, the authors employed a parameter that required learning to achieve a balance between conversation generation objective and quotation recommendation objective. \citet{yang2022improving} proposed a conversational recommendation framework that contain a generation module and a recommendation module. The overall objectives were designed to balance these two modules with a parameter learned through a fine-tuning process. However, improper optimization can lead to other problems as pointed out by \citet{deng2022unified} that "Error Propagation" may occur when solving multiple tasks in sequential order, leading to a decrease in performance with the sequential completion of each task. Although some potential solutions to this issue \citep{deng2022unified, li2022self, geng2022improving} were suggested, further verification is still needed. 

\noindent\textbf{Multiple Choices of PLM as Recommendation Bases.}
With the advances in variational PLMs, including ChatGPT, and their success in various downstream tasks, researchers have started exploring the potential of ChatGPT in conversational recommendation tasks. For example, \citet{liu2023chatgpt} and \citet{gao2023chat} have investigated the ability of GPT-3/GPT-3.5-based ChatGPT in zero-shot scenarios, using human-designed prompts to assess its performance in rating prediction, sequential recommendation, direct recommendation, and explanation generation. However, these studies are just initial explorations, and more extensive research is required on different recommendation tasks based on various pre-trained language models. This includes prompt design and performance evaluation in diverse domains. Moreover, recent LMRS studies have yet to explore instruction tuning, which could be a promising direction for future research.

\noindent\textbf{Privacy issue.}
The study conducted by \citet{yuan2020parameter} revealed that pre-trained models can infer user profiles (such as gender, age, and marital status) based on learned user representations, which raises concerns about privacy protection. The pre-training process is often performed on large-scale web-crawled corpus without fine-grained filtering, which may expose users' sensitive information. Therefore, developing LMRS that strike a balance between privacy and high-performance recommendation algorithms remains an open issue.

\section*{Acknowledgments}
We sincerely thank the action editor and the anonymous reviewers for their detailed feedback and helpful suggestions. This work is supported by the Research Council of Norway under Grant No. 309834.

\bibliography{tacl2021}

\begin{thebibliography}{72}
\expandafter\ifx\csname natexlab\endcsname\relax\def\natexlab#1{#1}\fi

\bibitem[{Bao et~al.(2020)Bao, Dong, Wei, Wang, Yang, Liu, Wang, Gao, Piao,
  Zhou, and Hon}]{bao2020unilmv2}
Hangbo Bao, Li~Dong, Furu Wei, Wenhui Wang, Nan Yang, Xiaodong Liu, Yu~Wang,
  Jianfeng Gao, Songhao Piao, Ming Zhou, and Hsiao-Wuen Hon. 2020.
\newblock {U}ni{LM}v2: Pseudo-masked language models for unified language model
  pre-training.
\newblock In \emph{Proceedings of the 37th International Conference on Machine
  Learning}, pages 642--652. PMLR.

\bibitem[{Brown et~al.(2020)Brown, Mann, Ryder, Subbiah, Kaplan, Dhariwal,
  Neelakantan, Shyam, Sastry, Askell, Agarwal, Herbert-Voss, Krueger, Henighan,
  Child, Ramesh, Ziegler, Wu, Winter, Hesse, Chen, Sigler, Litwin, Gray, Chess,
  Clark, Berner, McCandlish, Radford, Sutskever, and
  Amodei}]{brown2020language}
Tom Brown, Benjamin Mann, Nick Ryder, Melanie Subbiah, Jared~D Kaplan, Prafulla
  Dhariwal, Arvind Neelakantan, Pranav Shyam, Girish Sastry, Amanda Askell,
  Sandhini Agarwal, Ariel Herbert-Voss, Gretchen Krueger, Tom Henighan, Rewon
  Child, Aditya Ramesh, Daniel Ziegler, Jeffrey Wu, Clemens Winter, Chris
  Hesse, Mark Chen, Eric Sigler, Mateusz Litwin, Scott Gray, Benjamin Chess,
  Jack Clark, Christopher Berner, Sam McCandlish, Alec Radford, Ilya Sutskever,
  and Dario Amodei. 2020.
\newblock Language models are few-shot learners.
\newblock In \emph{Advances in Neural Information Processing Systems},
  volume~33, pages 1877--1901. Curran Associates, Inc.

\bibitem[{Chen et~al.(2023)Chen, Yuan, Yang, He, Li, and Yang}]{chen2021user}
Lei Chen, Fajie Yuan, Jiaxi Yang, Xiangnan He, Chengming Li, and Min Yang.
  2023.
\newblock \href {https://doi.org/10.1109/TKDE.2021.3119619} {User-specific
  adaptive fine-tuning for cross-domain recommendations}.
\newblock \emph{IEEE Transactions on Knowledge and Data Engineering},
  35(3):3239--3252.

\bibitem[{Deng et~al.(2023)Deng, Zhang, Xu, Lei, Chua, and
  Lam}]{deng2022unified}
Yang Deng, Wenxuan Zhang, Weiwen Xu, Wenqiang Lei, Tat-Seng Chua, and Wai Lam.
  2023.
\newblock \href {https://doi.org/10.1145/3570640} {A unified multi-task
  learning framework for multi-goal conversational recommender systems}.
\newblock \emph{ACM Transactions on Information Systems}, 41(3):1--25.

\bibitem[{Devlin et~al.(2019)Devlin, Chang, Lee, and
  Toutanova}]{devlin2019bert}
Jacob Devlin, Ming-Wei Chang, Kenton Lee, and Kristina Toutanova. 2019.
\newblock \href {https://doi.org/10.18653/v1/N19-1423} {{BERT}: Pre-training of
  deep bidirectional transformers for language understanding}.
\newblock In \emph{Proceedings of the 2019 Conference of the North {A}merican
  Chapter of the Association for Computational Linguistics: Human Language
  Technologies, Volume 1 (Long and Short Papers)}, pages 4171--4186,
  Minneapolis, Minnesota. Association for Computational Linguistics.

\bibitem[{Erhan et~al.(2010)Erhan, Courville, Bengio, and
  Vincent}]{erhan2010does}
Dumitru Erhan, Aaron Courville, Yoshua Bengio, and Pascal Vincent. 2010.
\newblock Why does unsupervised pre-training help deep learning?
\newblock In \emph{Proceedings of the Thirteenth International Conference on
  Artificial Intelligence and Statistics}, pages 201--208, Chia Laguna Resort,
  Sardinia, Italy. PMLR.

\bibitem[{Gao et~al.(2023)Gao, Sheng, Xiang, Xiong, Wang, and
  Zhang}]{gao2023chat}
Yunfan Gao, Tao Sheng, Youlin Xiang, Yun Xiong, Haofen Wang, and Jiawei Zhang.
  2023.
\newblock Chat-{REC}: Towards interactive and explainable {LLM}s-augmented
  recommender system.
\newblock \emph{arXiv preprint arXiv:2303.14524v2}.

\bibitem[{Geng et~al.(2022{\natexlab{a}})Geng, Fu, Ge, Li, de~Melo, and
  Zhang}]{geng2022improving}
Shijie Geng, Zuohui Fu, Yingqiang Ge, Lei Li, Gerard de~Melo, and Yongfeng
  Zhang. 2022{\natexlab{a}}.
\newblock \href {https://doi.org/10.18653/v1/2022.acl-long.20} {Improving
  personalized explanation generation through visualization}.
\newblock In \emph{Proceedings of the 60th Annual Meeting of the Association
  for Computational Linguistics (Volume 1: Long Papers)}, pages 244--255.
  Association for Computational Linguistics.

\bibitem[{Geng et~al.(2022{\natexlab{b}})Geng, Fu, Tan, Ge, De~Melo, and
  Zhang}]{geng2022path}
Shijie Geng, Zuohui Fu, Juntao Tan, Yingqiang Ge, Gerard De~Melo, and Yongfeng
  Zhang. 2022{\natexlab{b}}.
\newblock \href {https://doi.org/10.1145/3485447.3511937} {Path language
  modeling over knowledge graphs for explainable recommendation}.
\newblock In \emph{Proceedings of the ACM Web Conference 2022}, pages 946--955.
  Association for Computing Machinery.

\bibitem[{Geng et~al.(2022{\natexlab{c}})Geng, Liu, Fu, Ge, and
  Zhang}]{geng2022recommendation}
Shijie Geng, Shuchang Liu, Zuohui Fu, Yingqiang Ge, and Yongfeng Zhang.
  2022{\natexlab{c}}.
\newblock \href {https://doi.org/10.1145/3523227.3546767} {Recommendation as
  language processing ({RLP}): A unified pretrain, personalized prompt \&
  predict paradigm ({P}5)}.
\newblock In \emph{Proceedings of the 16th ACM Conference on Recommender
  Systems}, pages 299--315. Association for Computing Machinery.

\bibitem[{Geng et~al.(2023)Geng, Tan, Liu, Fu, and Zhang}]{geng2023vip5}
Shijie Geng, Juntao Tan, Shuchang Liu, Zuohui Fu, and Yongfeng Zhang. 2023.
\newblock {VIP}5: Towards multimodal foundation models for recommendation.
\newblock \emph{arXiv preprint arXiv:2305.14302v1}.

\bibitem[{Guo et~al.(2023)Guo, Wang, Wang, Zhu, and Yin}]{Guo2023AutomatedPF}
Lei Guo, Chunxiao Wang, Xinhua Wang, Lei Zhu, and Hongzhi Yin. 2023.
\newblock Automated prompting for non-overlapping cross-domain sequential
  recommendation.
\newblock \emph{arXiv preprint arXiv:2304.04218v1}.

\bibitem[{Hada and Shevade(2021)}]{hada2021rexplug}
Deepesh~V Hada and Shirish~K Shevade. 2021.
\newblock \href {https://doi.org/10.1145/3404835.3462939} {Re{XPl}ug:
  Explainable recommendation using plug-and-play language model}.
\newblock In \emph{Proceedings of the 44th International ACM SIGIR Conference
  on Research and Development in Information Retrieval}, pages 81--91.
  Association for Computing Machinery.

\bibitem[{He et~al.(2022)He, Xu, Yang, Han, Yang, and Lo}]{he2022ptm4tag}
Junda He, Bowen Xu, Zhou Yang, DongGyun Han, Chengran Yang, and David Lo. 2022.
\newblock \href {https://doi.org/10.1145/3524610.3527897} {{PTM}4{T}ag:
  Sharpening tag recommendation of stack overflow posts with pre-trained
  models}.
\newblock In \emph{Proceedings of the 30th IEEE/ACM International Conference on
  Program Comprehension}, pages 1--11. Association for Computing Machinery.

\bibitem[{Hou et~al.(2023)Hou, He, McAuley, and Zhao}]{Hou2022LearningVI}
Yupeng Hou, Zhankui He, Julian McAuley, and Wayne~Xin Zhao. 2023.
\newblock \href {https://doi.org/10.1145/3543507.3583434} {Learning
  vector-quantized item representation for transferable sequential
  recommenders}.
\newblock page 1162–1171. Association for Computing Machinery.

\bibitem[{Hou et~al.(2022)Hou, Mu, Zhao, Li, Ding, and Wen}]{hou2022towards}
Yupeng Hou, Shanlei Mu, Wayne~Xin Zhao, Yaliang Li, Bolin Ding, and Ji-Rong
  Wen. 2022.
\newblock \href {https://doi.org/10.1145/3534678.3539381} {Towards universal
  sequence representation learning for recommender systems}.
\newblock In \emph{Proceedings of the 28th ACM SIGKDD Conference on Knowledge
  Discovery and Data Mining}, pages 585--593. Association for Computing
  Machinery.

\bibitem[{JIANG et~al.(2022)JIANG, Xue, Zhang, Liu, Zhu, and
  Hao}]{jiang2022learning}
Caigao JIANG, Siqiao Xue, James~Y Zhang, Lingyue Liu, Zhibo Zhu, and Hongyan
  Hao. 2022.
\newblock Learning large-scale universal user representation with sparse
  mixture of experts.
\newblock In \emph{First Workshop on Pre-training: Perspectives, Pitfalls, and
  Paths Forward at ICML 2022}.

\bibitem[{Kang et~al.(2021)Kang, Wang, Zhang, Chen, and You}]{kang2021apirecx}
Yuning Kang, Zan Wang, Hongyu Zhang, Junjie Chen, and Hanmo You. 2021.
\newblock \href {https://doi.org/10.18653/v1/2021.emnlp-main.275} {{APIR}ec{X}:
  Cross-library {API} recommendation via pre-trained language model}.
\newblock In \emph{Proceedings of the 2021 Conference on Empirical Methods in
  Natural Language Processing}, pages 3425--3436. Association for Computational
  Linguistics.

\bibitem[{Li et~al.(2023{\natexlab{a}})Li, Zhang, Wang, Xiong, Lu, and
  Medioni}]{Li2023GPT4RecAG}
Jinming Li, Wentao Zhang, Tian Wang, Guanglei Xiong, Alan Lu, and Gérard
  Medioni. 2023{\natexlab{a}}.
\newblock \href
  {https://www.amazon.science/publications/gpt4rec-a-generative-framework-for-personalized-recommendation-and-user-interests-interpretation}
  {{GPT}4{R}ec: A generative framework for personalized recommendation and user
  interests interpretation}.
\newblock In \emph{SIGIR 2023 Workshop on eCommerce}.

\bibitem[{Li et~al.(2021)Li, Zhang, and Chen}]{li2021personalized}
Lei Li, Yongfeng Zhang, and Li~Chen. 2021.
\newblock \href {https://doi.org/10.18653/v1/2021.acl-long.383} {Personalized
  transformer for explainable recommendation}.
\newblock In \emph{Proceedings of the 59th Annual Meeting of the Association
  for Computational Linguistics and the 11th International Joint Conference on
  Natural Language Processing (Volume 1: Long Papers)}, pages 4947--4957.
  Association for Computational Linguistics.

\bibitem[{Li et~al.(2023{\natexlab{b}})Li, Zhang, and Chen}]{TOIS23-PEPLER}
Lei Li, Yongfeng Zhang, and Li~Chen. 2023{\natexlab{b}}.
\newblock \href {https://doi.org/10.1145/3580488} {Personalized prompt learning
  for explainable recommendation}.
\newblock \emph{ACM Transactions on Information Systems}, 41(4):1--26.

\bibitem[{Li et~al.(2022)Li, Xie, Zhu, Zhuang, Tang, Zhao, and He}]{li2022self}
Shuokai Li, Ruobing Xie, Yongchun Zhu, Fuzhen Zhuang, Zhenwei Tang, Wayne~Xin
  Zhao, and Qing He. 2022.
\newblock \href {https://doi.org/https://doi.org/10.1016/j.ipm.2022.103067}
  {Self-supervised learning for conversational recommendation}.
\newblock \emph{Information Processing \& Management}, 59(6):103067.

\bibitem[{Liu et~al.(2023{\natexlab{a}})Liu, Liu, Lv, Zhou, and
  Zhang}]{liu2023chatgpt}
Junling Liu, Chao Liu, Renjie Lv, Kang Zhou, and Yan Zhang. 2023{\natexlab{a}}.
\newblock Is {ChatGPT} a good recommender? {A} preliminary study.
\newblock \emph{arXiv preprint arXiv:2304.10149v2}.

\bibitem[{Liu et~al.(2023{\natexlab{b}})Liu, Yuan, Fu, Jiang, Hayashi, and
  Neubig}]{liu2023pre}
Pengfei Liu, Weizhe Yuan, Jinlan Fu, Zhengbao Jiang, Hiroaki Hayashi, and
  Graham Neubig. 2023{\natexlab{b}}.
\newblock \href {https://doi.org/10.1145/3560815} {Pre-train, prompt, and
  predict: A systematic survey of prompting methods in natural language
  processing}.
\newblock \emph{ACM Computing Surveys}, 55(9):1--35.

\bibitem[{Liu et~al.(2022)Liu, Zhu, Dai, and Wu}]{liu2022boosting}
Qijiong Liu, Jieming Zhu, Quanyu Dai, and Xiaoming Wu. 2022.
\newblock Boosting deep {CTR} prediction with a plug-and-play pre-trainer for
  news recommendation.
\newblock In \emph{Proceedings of the 29th International Conference on
  Computational Linguistics}, pages 2823--2833. International Committee on
  Computational Linguistics.

\bibitem[{Liu et~al.(2023{\natexlab{c}})Liu, Meng, Macdonald, and
  Ounis}]{liu2022graph}
Siwei Liu, Zaiqiao Meng, Craig Macdonald, and Iadh Ounis. 2023{\natexlab{c}}.
\newblock \href {https://doi.org/10.1145/3568953} {Graph neural pre-training
  for recommendation with side information}.
\newblock \emph{ACM Transactions on Information Systems}, 41(3):1--28.

\bibitem[{Liu et~al.(2019)Liu, Ott, Goyal, Du, Joshi, Chen, Levy, Lewis,
  Zettlemoyer, and Stoyanov}]{liu2019roberta}
Yinhan Liu, Myle Ott, Naman Goyal, Jingfei Du, Mandar Joshi, Danqi Chen, Omer
  Levy, Mike Lewis, Luke Zettlemoyer, and Veselin Stoyanov. 2019.
\newblock {R}o{BERT}a: A robustly optimized {BERT} pretraining approach.
\newblock \emph{arXiv preprint arXiv:1907.11692v1}.

\bibitem[{Liu et~al.(2023{\natexlab{d}})Liu, Jin, Pan, Zhou, Zheng, Xia, and
  Yu}]{liu2022graphsurvey}
Yixin Liu, Ming Jin, Shirui Pan, Chuan Zhou, Yu~Zheng, Feng Xia, and Philip Yu.
  2023{\natexlab{d}}.
\newblock \href {https://doi.org/10.1109/TKDE.2022.3172903} {Graph
  self-supervised learning: A survey}.
\newblock \emph{IEEE Transactions on Knowledge and Data Engineering},
  35(6):5879--5900.

\bibitem[{Liu et~al.(2021)Liu, Yang, Lei, Wang, Tang, Zhang, Sun, and
  Miao}]{liu2021pre}
Yong Liu, Susen Yang, Chenyi Lei, Guoxin Wang, Haihong Tang, Juyong Zhang,
  Aixin Sun, and Chunyan Miao. 2021.
\newblock \href {https://doi.org/10.1145/3474085.3475709} {Pre-training graph
  transformer with multimodal side information for recommendation}.
\newblock In \emph{Proceedings of the 29th ACM International Conference on
  Multimedia}, pages 2853--2861. Association for Computing Machinery.

\bibitem[{Long et~al.(2022)Long, Cao, Han, and Yang}]{ijcai2022p773}
Siqu Long, Feiqi Cao, Soyeon~Caren Han, and Haiqin Yang. 2022.
\newblock \href {https://doi.org/10.24963/ijcai.2022/773} {Vision-and-language
  pretrained models: A survey}.
\newblock In \emph{Proceedings of the Thirty-First International Joint
  Conference on Artificial Intelligence, {IJCAI-22}}, pages 5530--5537.
  International Joint Conferences on Artificial Intelligence Organization.
\newblock Survey Track.

\bibitem[{Long et~al.(2023)Long, Hui, Yuan, Huang, Li, and
  Wang}]{long2023multimodal}
Yuxing Long, Binyuan Hui, Caixia Yuan, Fei Huang, Yongbin Li, and Xiaojie Wang.
  2023.
\newblock \href {https://doi.org/10.18653/v1/2023.findings-acl.217} {Multimodal
  recommendation dialog with subjective preference: A new challenge and
  benchmark}.
\newblock In \emph{Findings of the Association for Computational Linguistics:
  ACL 2023}, pages 3515--3533, Toronto, Canada. Association for Computational
  Linguistics.

\bibitem[{Malkiel et~al.(2020)Malkiel, Barkan, Caciularu, Razin, Katz, and
  Koenigstein}]{malkiel2020recobert}
Itzik Malkiel, Oren Barkan, Avi Caciularu, Noam Razin, Ori Katz, and Noam
  Koenigstein. 2020.
\newblock \href {https://doi.org/10.18653/v1/2020.findings-emnlp.154}
  {{R}eco{BERT}: A catalog language model for text-based recommendations}.
\newblock In \emph{Findings of the Association for Computational Linguistics:
  EMNLP 2020}, pages 1704--1714. Association for Computational Linguistics.

\bibitem[{McKee et~al.(2023)McKee, Salamon, Sivic, and
  Russell}]{McKee_2023_CVPR}
Daniel McKee, Justin Salamon, Josef Sivic, and Bryan Russell. 2023.
\newblock \href {https://doi.org/10.1109/CVPR52729.2023.01420} {Language-guided
  music recommendation for video via prompt analogies}.
\newblock In \emph{Proceedings of the IEEE/CVF Conference on Computer Vision
  and Pattern Recognition (CVPR)}, pages 14784--14793. IEEE Computer Society.

\bibitem[{Penha and Hauff(2020)}]{penha2020does}
Gustavo Penha and Claudia Hauff. 2020.
\newblock \href {https://doi.org/10.1145/3383313.3412249} {What does {BERT}
  know about books, movies and music? {P}robing {BERT} for conversational
  recommendation}.
\newblock In \emph{Proceedings of the 14th ACM Conference on Recommender
  Systems}, pages 388--397. Association for Computing Machinery.

\bibitem[{Qin and Eisner(2021)}]{qin2021learning}
Guanghui Qin and Jason Eisner. 2021.
\newblock \href {https://doi.org/10.18653/v1/2021.naacl-main.410} {Learning how
  to ask: Querying {LM}s with mixtures of soft prompts}.
\newblock In \emph{Proceedings of the 2021 Conference of the North American
  Chapter of the Association for Computational Linguistics: Human Language
  Technologies}, pages 5203--5212. Association for Computational Linguistics.

\bibitem[{Qiu et~al.(2020)Qiu, Sun, Xu, Shao, Dai, and Huang}]{qiu2020pre}
Xipeng Qiu, Tianxiang Sun, Yige Xu, Yunfan Shao, Ning Dai, and Xuanjing Huang.
  2020.
\newblock \href {https://doi.org/10.1007/s11431-020-1647-3} {Pre-trained models
  for natural language processing: A survey}.
\newblock \emph{Science China Technological Sciences}, 63(10):1872--1897.

\bibitem[{Qiu et~al.(2021)Qiu, Wu, Gao, and Fan}]{qiu2021u}
Zhaopeng Qiu, Xian Wu, Jingyue Gao, and Wei Fan. 2021.
\newblock \href {https://doi.org/10.1609/aaai.v35i5.16557} {U-{BERT}:
  Pre-training user representations for improved recommendation}.
\newblock In \emph{Proceedings of the AAAI Conference on Artificial
  Intelligence}, volume~35, pages 4320--4327.

\bibitem[{Sankar et~al.(2021)Sankar, Wang, Krishnan, and
  Sundaram}]{sankar2021protocf}
Aravind Sankar, Junting Wang, Adit Krishnan, and Hari Sundaram. 2021.
\newblock \href {https://doi.org/10.1145/3460231.3474268} {Proto{CF}:
  Prototypical collaborative filtering for few-shot recommendation}.
\newblock In \emph{Proceedings of the 15th ACM Conference on Recommender
  Systems}, pages 166--175. Association for Computing Machinery.

\bibitem[{Sarkar et~al.(2022)Sarkar, Bodla, Vasileva, Lin, Beniwal, Lu, and
  Medioni}]{sarkar2022outfittransformer}
Rohan Sarkar, Navaneeth Bodla, Mariya Vasileva, Yen-Liang Lin, Anurag Beniwal,
  Alan Lu, and Gerard Medioni. 2022.
\newblock \href {https://doi.org/10.1109/CVPRW56347.2022.00249}
  {Outfit{T}ransformer: Outfit representations for fashion recommendation}.
\newblock In \emph{2022 IEEE/CVF Conference on Computer Vision and Pattern
  Recognition Workshops (CVPRW)}, pages 2262--2266.

\bibitem[{Shang et~al.(2019)Shang, Ma, Xiao, and Sun}]{ijcai2019p825}
Junyuan Shang, Tengfei Ma, Cao Xiao, and Jimeng Sun. 2019.
\newblock \href {https://doi.org/10.24963/ijcai.2019/825} {Pre-training of
  graph augmented transformers for medication recommendation}.
\newblock In \emph{Proceedings of the Twenty-Eighth International Joint
  Conference on Artificial Intelligence, {IJCAI-19}}, pages 5953--5959.
  International Joint Conferences on Artificial Intelligence.

\bibitem[{Shin et~al.(2023)Shin, Kwak, Kim, Ramstr{\"o}m, Jeong, Ha, and
  Kim}]{shin2023scaling}
Kyuyong Shin, Hanock Kwak, Su~Young Kim, Max~Nihl{\'e}n Ramstr{\"o}m, Jisu
  Jeong, Jung-Woo Ha, and Kyung-Min Kim. 2023.
\newblock \href {https://doi.org/10.1609/aaai.v37i4.25582} {Scaling law for
  recommendation models: Towards general-purpose user representations}.
\newblock In \emph{Proceedings of the AAAI Conference on Artificial
  Intelligence}, volume~37, pages 4596--4604.

\bibitem[{Sileo et~al.(2022)Sileo, Vossen, and Raymaekers}]{sileo2022zero}
Damien Sileo, Wout Vossen, and Robbe Raymaekers. 2022.
\newblock \href {https://doi.org/10.1007/978-3-030-99739-7_26} {Zero-shot
  recommendation as language modeling}.
\newblock In \emph{Advances in Information Retrieval: 44th European Conference
  on IR Research, ECIR 2022, Stavanger, Norway, April 10--14, 2022,
  Proceedings, Part II}, pages 223--230. Springer International Publishing.

\bibitem[{de~Souza Pereira~Moreira et~al.(2021)de~Souza Pereira~Moreira, Rabhi,
  Lee, Ak, and Oldridge}]{de2021transformers4rec}
Gabriel de~Souza Pereira~Moreira, Sara Rabhi, Jeong~Min Lee, Ronay Ak, and Even
  Oldridge. 2021.
\newblock \href {https://doi.org/10.1145/3460231.3474255} {Transformers4{R}ec:
  Bridging the gap between nlp and sequential/session-based recommendation}.
\newblock In \emph{Proceedings of the 15th ACM Conference on Recommender
  Systems}, pages 143--153. Association for Computing Machinery.

\bibitem[{Sun et~al.(2019)Sun, Liu, Wu, Pei, Lin, Ou, and
  Jiang}]{sun2019bert4rec}
Fei Sun, Jun Liu, Jian Wu, Changhua Pei, Xiao Lin, Wenwu Ou, and Peng Jiang.
  2019.
\newblock \href {https://doi.org/10.1145/3357384.3357895} {{BERT}4{R}ec:
  Sequential recommendation with bidirectional encoder representations from
  transformer}.
\newblock In \emph{Proceedings of the 28th ACM International Conference on
  Information and Knowledge Management}, pages 1441--1450. Association for
  Computing Machinery.

\bibitem[{Wang et~al.(2023{\natexlab{a}})Wang, Zhou, Zhao, Wang, and
  Wen}]{wang2023curriculum}
Hui Wang, Kun Zhou, Xin Zhao, Jingyuan Wang, and Ji-Rong Wen.
  2023{\natexlab{a}}.
\newblock \href {https://doi.org/10.1145/3528667} {Curriculum pre-training
  heterogeneous subgraph transformer for top-n recommendation}.
\newblock \emph{ACM Transactions on Information Systems}, 41(1):1--28.

\bibitem[{Wang et~al.(2022{\natexlab{a}})Wang, Hu, Sha, Xu, Jiang, and
  Wong}]{wang2022recindial}
Lingzhi Wang, Huang Hu, Lei Sha, Can Xu, Daxin Jiang, and Kam-Fai Wong.
  2022{\natexlab{a}}.
\newblock {R}ec{I}n{D}ial: A unified framework for conversational
  recommendation with pretrained language models.
\newblock In \emph{Proceedings of the 2nd Conference of the Asia-Pacific
  Chapter of the Association for Computational Linguistics and the 12th
  International Joint Conference on Natural Language Processing (Volume 1: Long
  Papers)}, pages 489--500. Association for Computational Linguistics.

\bibitem[{Wang et~al.(2022{\natexlab{b}})Wang, Zeng, and
  Wong}]{wang2022learning}
Lingzhi Wang, Xingshan Zeng, and Kam-Fai Wong. 2022{\natexlab{b}}.
\newblock \href {https://doi.org/10.18653/v1/2022.findings-emnlp.225} {Learning
  when and what to quote: A quotation recommender system with mutual promotion
  of recommendation and generation}.
\newblock In \emph{Findings of the Association for Computational Linguistics:
  EMNLP 2022}, pages 3094--3105. Association for Computational Linguistics.

\bibitem[{Wang et~al.(2023{\natexlab{b}})Wang, Zeng, and
  Wong}]{wang2023quotation}
Lingzhi Wang, Xingshan Zeng, and Kam-Fai Wong. 2023{\natexlab{b}}.
\newblock \href {https://doi.org/10.1145/3594633} {Quotation recommendation for
  multi-party online conversations based on semantic and topic fusion}.
\newblock \emph{ACM Transactions on Information Systems}.

\bibitem[{Wang et~al.(2022{\natexlab{c}})Wang, Zhou, Wen, and
  Zhao}]{wang2022towards}
Xiaolei Wang, Kun Zhou, Ji-Rong Wen, and Wayne~Xin Zhao. 2022{\natexlab{c}}.
\newblock \href {https://doi.org/10.1145/3534678.3539382} {Towards unified
  conversational recommender systems via knowledge-enhanced prompt learning}.
\newblock In \emph{Proceedings of the 28th ACM SIGKDD Conference on Knowledge
  Discovery and Data Mining}, page 1929–1937. Association for Computing
  Machinery.

\bibitem[{Wu et~al.(2021)Wu, Wu, Qi, and Huang}]{wu2021empowering}
Chuhan Wu, Fangzhao Wu, Tao Qi, and Yongfeng Huang. 2021.
\newblock \href {https://doi.org/10.1145/3404835.3463069} {Empowering news
  recommendation with pre-trained language models}.
\newblock In \emph{Proceedings of the 44th International ACM SIGIR Conference
  on Research and Development in Information Retrieval}, page 1652–1656.
  Association for Computing Machinery.

\bibitem[{Wu et~al.(2022{\natexlab{a}})Wu, Wu, Qi, Zhang, Huang, and
  Xu}]{wu2022mm}
Chuhan Wu, Fangzhao Wu, Tao Qi, Chao Zhang, Yongfeng Huang, and Tong Xu.
  2022{\natexlab{a}}.
\newblock \href {https://doi.org/10.1145/3477495.3531896} {{MM-R}ec:
  Visiolinguistic model empowered multimodal news recommendation}.
\newblock In \emph{Proceedings of the 45th International ACM SIGIR Conference
  on Research and Development in Information Retrieval}, page 2560–2564.
  Association for Computing Machinery.

\bibitem[{Wu et~al.(2022{\natexlab{b}})Wu, Xie, Zhu, Zhuang, Zhang, Lin, and
  He}]{wu2022personalized}
Yiqing Wu, Ruobing Xie, Yongchun Zhu, Fuzhen Zhuang, Xu~Zhang, Leyu Lin, and
  Qing He. 2022{\natexlab{b}}.
\newblock Personalized prompts for sequential recommendation.
\newblock \emph{arXiv preprint arXiv:2205.09666v2}.

\bibitem[{Xiao et~al.(2021)Xiao, Xie, Yao, Liu, Sun, Zhang, and
  Lin}]{xiao2021uprec}
Chaojun Xiao, Ruobing Xie, Yuan Yao, Zhiyuan Liu, Maosong Sun, Xu~Zhang, and
  Leyu Lin. 2021.
\newblock {UPR}ec: User-aware pre-training for recommender systems.
\newblock \emph{arXiv preprint arXiv:2102.10989v1}.

\bibitem[{Xiao et~al.(2022)Xiao, Liu, Shao, Di, Middha, Wu, and
  Xie}]{xiao2022training}
Shitao Xiao, Zheng Liu, Yingxia Shao, Tao Di, Bhuvan Middha, Fangzhao Wu, and
  Xing Xie. 2022.
\newblock \href {https://doi.org/10.1145/3534678.3539120} {Training large-scale
  news recommenders with pretrained language models in the loop}.
\newblock In \emph{Proceedings of the 28th ACM SIGKDD Conference on Knowledge
  Discovery and Data Mining}, page 4215–4225. Association for Computing
  Machinery.

\bibitem[{Xie et~al.(2023)Xie, Singh, McAuley, and Majumder}]{xie2022factual}
Zhouhang Xie, Sameer Singh, Julian McAuley, and Bodhisattwa~Prasad Majumder.
  2023.
\newblock \href {https://doi.org/10.1609/aaai.v37i11.26618} {Factual and
  informative review generation for explainable recommendation}.
\newblock \emph{Proceedings of the AAAI Conference on Artificial Intelligence},
  pages 13816--13824.

\bibitem[{Xin et~al.(2022)Xin, Pimentel, Karatzoglou, Ren, Christakopoulou, and
  Ren}]{xin2022rethinking}
Xin Xin, Tiago Pimentel, Alexandros Karatzoglou, Pengjie Ren, Konstantina
  Christakopoulou, and Zhaochun Ren. 2022.
\newblock \href {https://doi.org/10.1145/3477495.3531714} {Rethinking
  reinforcement learning for recommendation: A prompt perspective}.
\newblock In \emph{Proceedings of the 45th International ACM SIGIR Conference
  on Research and Development in Information Retrieval}, page 1347–1357.
  Association for Computing Machinery.

\bibitem[{Yang et~al.(2022{\natexlab{a}})Yang, Han, Li, Zuo, and
  Yu}]{yang2022improving}
Bowen Yang, Cong Han, Yu~Li, Lei Zuo, and Zhou Yu. 2022{\natexlab{a}}.
\newblock \href {https://doi.org/10.18653/v1/2022.findings-naacl.4} {Improving
  conversational recommendation systems{'} quality with context-aware item
  meta-information}.
\newblock In \emph{Findings of the Association for Computational Linguistics:
  NAACL 2022}, pages 38--48. Association for Computational Linguistics.

\bibitem[{Yang et~al.(2022{\natexlab{b}})Yang, Kim, Kim, and
  Park}]{yang-etal-2022-gram}
Yoonseok Yang, Kyu~Seok Kim, Minsam Kim, and Juneyoung Park.
  2022{\natexlab{b}}.
\newblock \href {https://doi.org/10.18653/v1/2022.naacl-main.61} {{GRAM}: Fast
  fine-tuning of pre-trained language models for content-based collaborative
  filtering}.
\newblock In \emph{Proceedings of the 2022 Conference of the North American
  Chapter of the Association for Computational Linguistics: Human Language
  Technologies}, pages 839--851. Association for Computational Linguistics.

\bibitem[{Yang et~al.(2019)Yang, Dai, Yang, Carbonell, Salakhutdinov, and
  Le}]{yang2019xlnet}
Zhilin Yang, Zihang Dai, Yiming Yang, Jaime Carbonell, Ruslan Salakhutdinov,
  and Quoc~V. Le. 2019.
\newblock \emph{{XLN}et: Generalized Autoregressive Pretraining for Language
  Understanding}. Curran Associates Inc.

\bibitem[{Yu et~al.(2023)Yu, Yin, Xia, Chen, Li, and Huang}]{yu2022self}
Junliang Yu, Hongzhi Yin, Xin Xia, Tong Chen, Jundong Li, and Zi~Huang. 2023.
\newblock \href {https://doi.org/10.1109/TKDE.2023.3282907} {Self-supervised
  learning for recommender systems: A survey}.
\newblock \emph{IEEE Transactions on Knowledge and Data Engineering}, pages
  1--20.

\bibitem[{Yu et~al.(2022)Yu, Wu, Wu, Yi, and Liu}]{yu2021tiny}
Yang Yu, Fangzhao Wu, Chuhan Wu, Jingwei Yi, and Qi~Liu. 2022.
\newblock \href {https://doi.org/10.18653/v1/2022.emnlp-main.368}
  {Tiny-{N}ews{R}ec: Effective and efficient {PLM}-based news recommendation}.
\newblock In \emph{Proceedings of the 2022 Conference on Empirical Methods in
  Natural Language Processing}, pages 5478--5489. Association for Computational
  Linguistics.

\bibitem[{Yuan et~al.(2020{\natexlab{a}})Yuan, He, Jiang, Guo, Xiong, Xu, and
  Xiong}]{yuan2020future}
Fajie Yuan, Xiangnan He, Haochuan Jiang, Guibing Guo, Jian Xiong, Zhezhao Xu,
  and Yilin Xiong. 2020{\natexlab{a}}.
\newblock \href {https://doi.org/10.1145/3366423.3380116} {Future data helps
  training: Modeling future contexts for session-based recommendation}.
\newblock In \emph{Proceedings of The Web Conference 2020}, page 303–313.
  Association for Computing Machinery.

\bibitem[{Yuan et~al.(2020{\natexlab{b}})Yuan, He, Karatzoglou, and
  Zhang}]{yuan2020parameter}
Fajie Yuan, Xiangnan He, Alexandros Karatzoglou, and Liguang Zhang.
  2020{\natexlab{b}}.
\newblock \href {https://doi.org/10.1145/3397271.3401156} {Parameter-efficient
  transfer from sequential behaviors for user modeling and recommendation}.
\newblock In \emph{Proceedings of the 43rd International ACM SIGIR Conference
  on Research and Development in Information Retrieval}, page 1469–1478.
  Association for Computing Machinery.

\bibitem[{Zeng et~al.(2021)Zeng, Xiao, Yao, Xie, Liu, Lin, Lin, and
  Sun}]{zeng2021knowledge}
Zheni Zeng, Chaojun Xiao, Yuan Yao, Ruobing Xie, Zhiyuan Liu, Fen Lin, Leyu
  Lin, and Maosong Sun. 2021.
\newblock \href {https://doi.org/10.3389/fdata.2021.602071} {Knowledge transfer
  via pre-training for recommendation: A review and prospect}.
\newblock \emph{Frontiers in big Data}, 4.

\bibitem[{Zhang et~al.(2021{\natexlab{a}})Zhang, Li, Jia, Wang, Zhu, Wang, and
  He}]{ijcai2021p462}
Qi~Zhang, Jingjie Li, Qinglin Jia, Chuyuan Wang, Jieming Zhu, Zhaowei Wang, and
  Xiuqiang He. 2021{\natexlab{a}}.
\newblock \href {https://doi.org/10.24963/ijcai.2021/462} {{UNBERT}: User-news
  matching bert for news recommendation}.
\newblock In \emph{Proceedings of the Thirtieth International Joint Conference
  on Artificial Intelligence, {IJCAI-21}}, pages 3356--3362. International
  Joint Conferences on Artificial Intelligence Organization.
\newblock Main Track.

\bibitem[{Zhang et~al.(2023)Zhang, Malkov, Florez, Park, McWilliams, Han, and
  El-Kishky}]{zhang2022twhin}
Xinyang Zhang, Yury Malkov, Omar Florez, Serim Park, Brian McWilliams, Jiawei
  Han, and Ahmed El-Kishky. 2023.
\newblock \href {https://doi.org/10.1145/3580305.3599921} {Tw{HIN-BERT}: A
  socially-enriched pre-trained language model for multilingual tweet
  representations at twitter}.
\newblock In \emph{Proceedings of the 29th ACM SIGKDD Conference on Knowledge
  Discovery and Data Mining}, page 5597–5607. Association for Computing
  Machinery.

\bibitem[{Zhang et~al.(2021{\natexlab{b}})Zhang, Ding, Shui, Ma, Zou, Deoras,
  and Wang}]{zhang2021language}
Yuhui Zhang, Hao Ding, Zeren Shui, Yifei Ma, James Zou, Anoop Deoras, and Hao
  Wang. 2021{\natexlab{b}}.
\newblock \href
  {https://www.amazon.science/publications/language-models-as-recommender-systems-evaluations-and-limitations}
  {Language models as recommender systems: Evaluations and limitations}.
\newblock In \emph{NeurIPS 2021 Workshop on I (Still) Can't Believe It's Not
  Better}.

\bibitem[{Zhang et~al.(2022)Zhang, Chan, Xu, Bian, Han, Deng, and
  Zheng}]{zhang2022keep}
Yujing Zhang, Zhangming Chan, Shuhao Xu, Weijie Bian, Shuguang Han, Hongbo
  Deng, and Bo~Zheng. 2022.
\newblock \href {https://doi.org/10.1145/3511808.3557106} {{KEEP}: An
  industrial pre-training framework for online recommendation via knowledge
  extraction and plugging}.
\newblock In \emph{Proceedings of the 31st ACM International Conference on
  Information \& Knowledge Management}, page 3684–3693. Association for
  Computing Machinery.

\bibitem[{Zhang and Wang(2023)}]{Zhang2023PromptLF}
Zizhuo Zhang and Bang Wang. 2023.
\newblock \href {https://doi.org/10.1145/3539618.3591752} {Prompt learning for
  news recommendation}.
\newblock In \emph{Proceedings of the 46th International ACM SIGIR Conference
  on Research and Development in Information Retrieval}, page 227–237.
  Association for Computing Machinery.

\bibitem[{Zhao(2022)}]{zhao2022resetbert4rec}
Qihang Zhao. 2022.
\newblock \href {https://doi.org/10.1145/3477495.3532054} {{RESETBERT4R}ec: A
  pre-training model integrating time and user historical behavior for
  sequential recommendation}.
\newblock In \emph{Proceedings of the 45th International ACM SIGIR Conference
  on Research and Development in Information Retrieval}, page 1812–1816.
  Association for Computing Machinery.

\bibitem[{Zheng et~al.(2022)Zheng, Yang, Wang, Yang, Li, Hu, and
  Hong}]{zheng2022spatial}
Jiayi Zheng, Ling Yang, Heyuan Wang, Cheng Yang, Yinghong Li, Xiaowei Hu, and
  Shenda Hong. 2022.
\newblock Spatial autoregressive coding for graph neural recommendation.
\newblock \emph{arXiv preprint arXiv:2205.09489v2}.

\bibitem[{Zhou et~al.(2020)Zhou, Wang, Zhao, Zhu, Wang, Zhang, Wang, and
  Wen}]{zhou2020s3}
Kun Zhou, Hui Wang, Wayne~Xin Zhao, Yutao Zhu, Sirui Wang, Fuzheng Zhang,
  Zhongyuan Wang, and Ji-Rong Wen. 2020.
\newblock \href {https://doi.org/10.1145/3340531.3411954} {S3-{R}ec:
  Self-supervised learning for sequential recommendation with mutual
  information maximization}.
\newblock In \emph{Proceedings of the 29th ACM International Conference on
  Information \& Knowledge Management}, page 1893–1902. Association for
  Computing Machinery.

\end{thebibliography}
\bibliographystyle{acl_natbib}

\iftaclpubformat

\onecolumn

\end{document}